\pdfoutput=1
\documentclass[aps,prl,twocolumn,superscriptaddress,showpacs,amsfonts,amsmath,amssymb]{revtex4-1}
\usepackage{graphicx}
\usepackage{amssymb}
\usepackage{natbib}
\usepackage{amsmath}
\usepackage{amsfonts}
\usepackage{bm}
\usepackage{subfigure}
\usepackage{comment}
\usepackage{verbatim}
\usepackage{color}
\usepackage{upgreek}

\begin{document}

\title{Iron based high transition temperature superconductors }

\author{Xianhui Chen}
\affiliation{Department of Physics, University of Science and Technology of China, Hefei, China}

\author{Pengcheng Dai}
\affiliation{Department of Physics and Astronomy, Rice University, Houston, Texas 77005, USA}
\affiliation{Institute of Physics, Chinese Academy of Sciences, Beijing 100190, China}

\author{Donglai Feng}
\affiliation{Department of Physics, Fudan University, Shanghai, China}

\author{Tao Xiang}
\affiliation{Institute of Physics, Chinese Academy of Sciences, Beijing 100190, China}
\affiliation{Collaborative Innovation Center of Quantum Matter, Beijing, China}

\author{Fu-Chun Zhang}
\affiliation{Department of Physics, Zhejiang University, Hangzhou, China}

\begin{abstract}
In a superconductor electrons form pairs and electric transport becomes dissipation-less at low temperatures. Recently discovered iron based superconductors have the highest superconducting transition temperature next to copper oxides. In this article, we review material aspects and physical properties of iron based superconductors.  We discuss the dependence of transition temperature on the crystal structure, the interplay between antiferromagnetism and superconductivity by examining neutron scattering experiments, and the electronic properties of these compounds obtained by angle resolved photoemission spectroscopy in link with some results from scanning tunneling microscopy/spectroscopy measurements.  Possible microscopic model for this class of compounds is discussed from a strong coupling point of view.
\end{abstract}

\maketitle

\section{Introduction}
Superconductivity is a remarkable macroscopic quantum phenomenon, which was discovered by Kamerlingh Onnes in 1911.  As temperature decreases to below a critical value, electric resistance of a superconductor vanishes and the magnetic field is repelled. Superconductors have many applications. As an example, magnetic resonance imaging has been widely used in medical facilities. Superconductors may be used to transport electricity without loss of energy. Conventional superconductivity is well explained by Bardeen-Cooper-Schrieffer (BCS) theory, which was established in 1957. In a superconducting (SC) state, two electrons with opposite momenta attract each other to form a bound pair.  The pairing mechanism in a conventional superconductor is due to couplings between electrons and phonons, which are quantum version of lattice vibration.  The transition temperatures ($T_cs$) are, however, very low, and usually well below 40 Kelvin (K).  The low transition temperature has greatly limited practical applications of superconductors.  It has been a dream to realize high $T_c$ or room temperature superconductors, which may revolutionally change the power transmission in the world.

There was a great excitement after the discovery of high $T_c$ SC cuprate by Bednorz and M$\rm \ddot{u}$ller in 1986, who reported $T_c $ well above 30 K in La$_{2-x}$Ca$_{x}$CuO$_4$  ~\cite{Bednorz}. The subsequent world-wide efforts in search of high $T_c$ SC cuprate raised the transition temperature beyond the liquid nitrogen temperature of 77K for the first time~\cite{Wu-Chu} and the highest $T_c$ at ambient pressure is 135K in Hg-based cuprates, which remains the record as of today. All the cuprates share a common structure element CuO$_2$ plane, where Cu atoms form a square lattice. The second class of high $T_c$ materials is iron based superconductors, which was discovered by Hosono and co-workers in early 2008~\cite{Hosono}, who reported $T_c =26$K in LaOFeAs with part of O atoms replaced by F atoms.  Soon after this  discovery, the transition temperature has been raised to above 40K at ambient pressure by substitution of different elements~\cite{ChenXH2008,WangNL2008,ZhaoZX2008,ZhaoZX2008-2}. The highest $T_c$ in bulk iron-based superconductors is 55K in SmO$_{1-x}$F$_x$FeAs reported by Ren et al. ~\cite{ZhaoZX2008}, and similarly in Gd$_{1-x}$Th$_x$FeAsO ~\cite{XuZA2008}. So far many families of iron-based superconductors have been discovered~\cite{WenHH2008,122,11,111,111-2}. Most recently, monolayer FeSe superconductivity on top of substrate SrTiO$_3$ has been reported~\cite{QKXueCPL}, and there is an indication that T$_c$ is likely higher than the bulk ones.

Study of iron based superconductors and their physical properties have been one of the major activities in condensed matter physics in the past several years. At the time when the iron based superconductor was discovered, scientists in the field of superconductivity were well prepared.  Several powerful new techniques such as angle resolved photoemission spectroscope (ARPES), scanning tunneling microscopy (STM), have been well developed during the course of studying high T$_c$ cuprates.  These techniques together with some more conventional techniques such as neutron scattering, nuclear magnetic resonance (NMR), optical conducting measurements, have been  applied to examine the properties of the new compounds. Iron based superconductivity shares many common features with the high $T_c$ cuprates. Both of them are unconventional superconductors in the sense that phonons  unlikely play any dominant role in their superconductivity.  Both are quasi-two dimensional, and their superconductivity is in the proximity  of antiferromagnetism. On the other hand, iron based superconductors have a number of distinct properties  from the cuprates.  The parent compounds of the SC cuprates are antiferromagnetic (AF) Mott insulators due to strong Coulomb repulsion, and the lightly doped superconductors are doped Mott insulators~\cite{Anderson87,Anderson04,LeeWenNagaosa,rice2012}. On the other hand, the parent compounds of iron based superconductors are semi-metallic.  Dynamic mean field calculations indicate that the iron based compounds are close to metal-insulator transition line but are at the metallic side~\cite{kotliar}.  In the curpates, the low energy physics is described by a single  band~\cite{ZR}, while in the iron-based compounds, there are multi orbitals involved. Despite over 25 years of study, some of the physics in the cuprates remain controversial. The investigation of iron based superconductors may help us to understand the unconventional superconductivity and also provide a new route for searching higher temperature superconductors.

The purpose of this article is to provide an overall picture of the
iron based superconductivity based on our present understanding.
Instead of giving a broad review to cover all the
experimental and theoretical developments in this field,  we will
discuss basic physical properties of the materials and the underlying
physics by examining limited experiments and theories.  We refer
the readers to several recent review
papers \cite{Johnston,Stewart,Dagotto,Dai2012,Greene-rev,Hirschfeld-rev,Chubukov-rev} for more complete
description of the field.   The rest of the article is
organized as follows. In section 2, we discuss materials aspect
of the compounds.  Antiferromagnetism and superconductivity will be
discussed in section 3.  In section 4, we discuss electronic
structure of iron-based materials largely based on ARPES and STM
experiments.  We briefly present our theoretical understanding of the
electronic structure and superconductivity in section 5.  The
article will end with a summary and perspective in section 6.
\newpage
\section{Materials and Crystal Structures}
\subsection{Material classification and crystal structures}

Substitution of magnetic Sm for non-magnetic La leads to a dramatic increase in $T_c$ from 26 K in LaFeAsO$_{1-x}$F$_x$ to 43 K in SmFeAsO$_{1-x}$F$_x$ \cite{Hosono,ChenXH2008}. This suggests that a higher T$_c$ is possibly realized in the layered oxypnictides. The achieved $T_c$ of 43 K in SmFeAsO$_{1-x}$F$_x$ is higher than the commonly believed the upper limit (40 K) of electron-phonon mediated superconductors, which gives compelling evidences for classifying layered iron-based superconductors as a family of unconventional superconductors. Subsequently, many new iron-based superconductors with diverse crystal structures were found and they can be categorized into several families according to their structural  features.
\begin{figure*} \includegraphics[scale=.8]{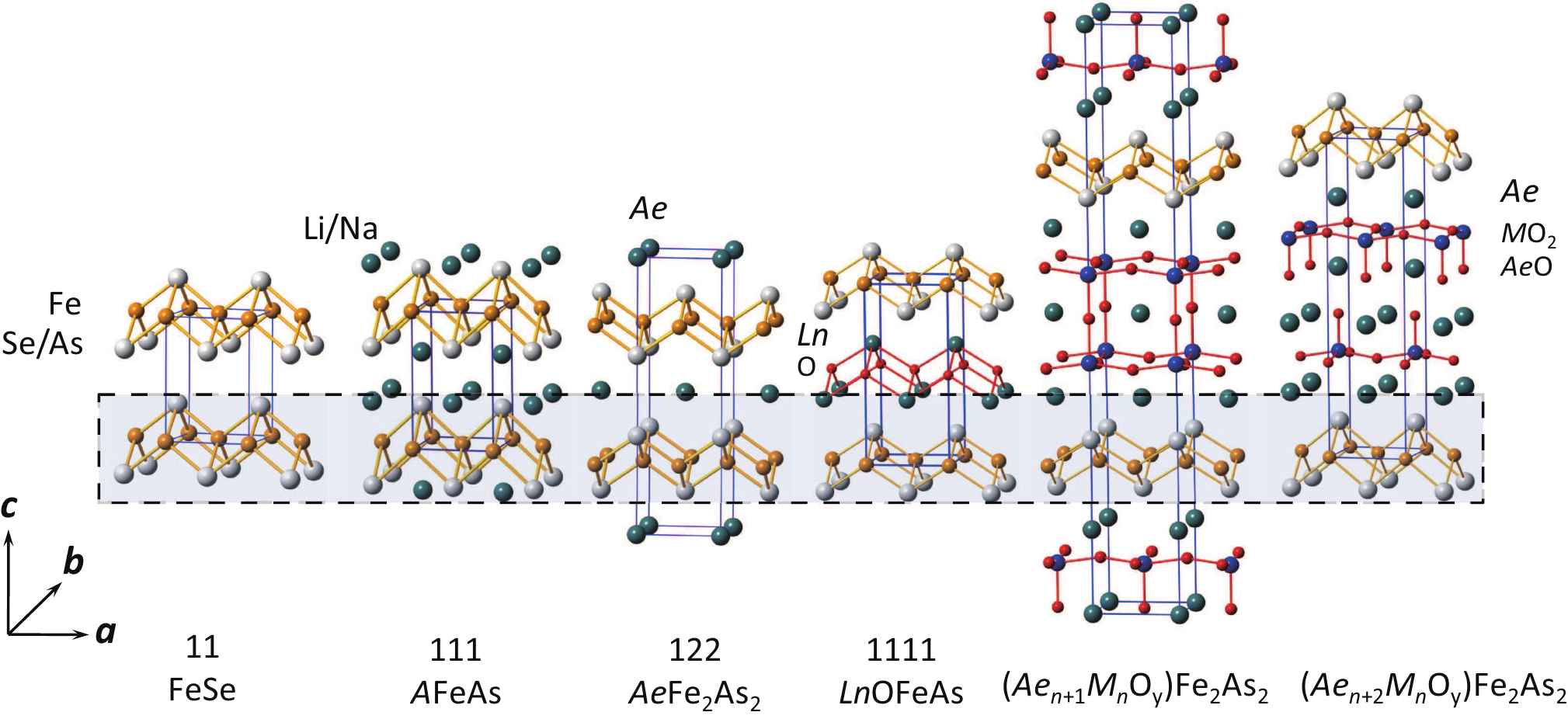}
\caption{
The schematic view of the crystal structures for several typical types of iron-based superconductors, in which A, Ae, Ln and M stand for alkali, alkali earth, lanthanide, and transition metal atoms.}
\label{fig:cxh1}
\end{figure*}

The iron-based superconductors share the common Fe$_2$X$_2$ (X=As and Se) layered structure unit, which possesses an anti-PbO-type (anti-litharge-type) atom arrangement. The Fe$_2$X$_2$ layers consist of edge-shared FeX$_{4/4}$ tetradedra, which has $\bar{4}$m2 site symmetry. In the Fe$_2$X$_2$ layers, X ions form a distorted tetrahedral arrangement around the Fe ions, giving rise to two distinct X-Fe-X bond angles with multiplicities of two and four which we refer to as $\alpha$ and $\beta$, respectively.

 FeSe ($T_c$ = 8 K) has the simplest structure among the known iron-based superconductors, which is called 11 phase \cite{11}. FeSe is formed by alternate stacking of the anti-PbO FeSe layers. In FeSe, the cations and anions occupy the opposite sites to Pb and O atoms of litharge, so that we call it anti-PbO or anti-litharge structure. FeSe adopts a space group of $P$4/$nmm$. The Fe$_2$Se$_2$ monolayer consists of flat Fe$_2$ square-net sandwiched by two Se monolayers. Consequently, each Fe atom is coordinated with four Se atoms to establish the edge-shared FeSe$_4$ tetrahedron, forming a 2D square-net Fe$_2$Se$_2$ monolayer. As shown in Fig. \ref{fig:cxh1}, these tetrahedral Fe$_2$X$_2$ layers can be separated by alkali and alkali-earth cations, LnO layers or perovskite-related oxydic slabs. Fig. \ref{fig:cxh1} also illustrates the crystal structures of AFeAs, AeFe$_2$As$_2$, LnOFeAs.  Their basic crystallographic data are listed in Table I. The simplest FeAs-based superconductor in structure is the AFeAs (A = Li and Na, called 111 phase) \cite{111,111-2,chen19}. AFeAs crystallizes in an anti-PbFCl-type structure, which adopts a Cu$_2$Sb (or Fe$_2$As) structure. AFeAs has the space group of \textit{P4/nmm} and each unit cell includes two chemical formula, that is 2A, 2Fe and 2As. Fe and As are arranged in anti-PbO-type layers with double Li/Na planes located between the layers in square-based pyramidal coordination by As.

With additional atoms added into the anti-PbFCl-type structure, we can achieve ZrCuSiAs-type 1111 superconductors. Up to now, the highest $T_c \sim$ 55 K in iron-based superconductors has been achieved in fluorine-doped or oxygen-deficient LnFeAsO compounds (Ln represents rare-earth metal atoms) \cite{ZhaoZX2008}, which are usually briefly written as 1111 phase. LnFeAsO compounds have a tetragonal layered structure at room temperature, with space group P4/nmm. The schematic view of their crystal structure is shown in Fig. \ref{fig:cxh1}. The earliest discovered 1111 compound with relative high $T_c$ is LaFeAsO \cite{Hosono}, with lattice constants at room temperature \emph{a} = 4.03268(1) \AA, \emph{c} = 8.74111(4) \AA. For these 1111 compounds, their structure consists of alternate stacking of FeAs layers and fluorite-type LnO layers. For LaFeAsO, the distance between the adjacent FeAs and LaO layers is 1.8 \AA. The lattice constants \emph{a} and \emph{c} decrease with reducing the ion radii of the rare-earth metals. With the decreasing radii of the rare-earth metal ions, the optimal $T_c$ first increases rapidly, reaching the highest $T_c$ (= 55 K) in the doped SmFeAsO system \cite{ZhaoZX2008,chen25}, and then decreases slightly with further reducing the radii of the rare-earth metal ions. Besides LnFeAsO systems, there are other types of 1111 FeAs-based compound, AeFFeAs (Ae = Ca, Sr and Ba) \cite{cxh26,cxh27} and CaHFeAs \cite{cxh28}. AeFFeAs (Ae =Ca, Sr and Ba) and CaHFeAs are also parent compounds of superconductors \cite{cxh29,cxh30,cxh31}. Very recently, a new 1111-type FeSe-derived superconductor, LiFeO$_2$Fe$_2$Se$_2$ with T$_c \approx$ 43 K, was synthesized by Lu \textit{et al} \cite{cxh32}.

 The other typical type of compounds, ThCr$_2$Si$_2$-type iron arsenides, possess only single layers of separating spacer atoms between Fe$_2$X$_2$ (so called 122 structure), which is adopted by AeFe$_2$As$_2$ (Ae = Ca, Sr, Ba, Eu, K etc.) \cite{122,cxh8,cxh9,cxh10} and A$_x$Fe$_{2-y}$Se$_2$ (A=K, Rb, Cs, Tl/K and Tl/Rb) \cite{cxh13,cxh14,cxh15,cxh16}. AeFe$_2$As$_2$ adopts body-centered tetragonal lattice and has space group of \textit{I4/mmm}. In FeAs-122, the highest T$_c\sim$ 49 K can be achieved in Pr-doped CaFe$_2$As$_2$ \cite{cxh11}. However, some recent reports revealed that such superconductivity with T$_c$ higher than 40 K should be ascribed to a new structural phase (Ca,Ln)FeAs$_2$ (Ln = La, Pr). Crystal structure of (Ca,Ln)FeAs$_2$ is derived from CaFe$_2$As$_2$, with shifting the adjacent FeAs layers along the 45$^\circ$ direction of ab-plane by half lattice length on the basis of CaFe$_2$As$_2$ and then intercalating one additional As-plane and one Ca-plane for every two CaFeAs blocks \cite{chen18}.
FeSe-derived superconductors A$_x$Fe$_{2-y}$Se$_2$ also crystallize in 122 structure, which have a T$_c$ of $\sim$ 30 K in crystals grown by the high-temperature melting method \cite{cxh13,cxh14,cxh15,cxh16} or higher than 40 K by co-intercalation of alkali atoms and  certain molecules (NH$_3$ or organic) by a low-temperature solution route \cite{cxh21,cxh22,cxh23,cxh24}.

According to the previous knowledge in the high-$T_c$ cuprates, superconductivity is closely related to the separating spacers between adjacent conducting layers. Therefore, compounds with complicated structures between FeAs layers were synthesized. Up to now, Ae$_{n+1}$M$_{n}$O$_y$Fe$_2$As$_2$ and Ae$_{n+2}$M$_{n}$O$_y$Fe$_2$As$_2$ (Ae = Ca, Sr, Ba; M = Sc, V, (Ti,Al), (Ti,Mg) and (Sc,Mg)) systems have been successfully synthesized, where $y$$\sim$3$n$-1 for the former and y$\sim$3n for the latter \cite{cxh33,cxh34,cxh35,cxh36}. Fig. \ref{fig:cxh1} show the crystal structures for the case of $n$=1. They all adopt tetragonal lattice. All Ae$_{n+1}$M$_{n}$O$_y$Fe$_2$As$_2$ compounds share the same space group of D$^{174}_h$-$I$4/$mmm$, while for  Ae$_{n+2}$M$_{n}$O$_y$Fe$_2$As$_2$, its space group is $P$4/$nmm$ for $n$ = 2 and 4, whereas $P$4$mm$ for $n$ = 3. For Ae$_{n+1}$M$_{n}$O$_y$Fe$_2$As$_2$, $n$ perovkite layers are sandwiched between adjacent FeAs layers, while for Ae$_{n+2}$M$_{n}$O$_y$Fe$_2$As$_2$ there are $n$ perovkite layers plus one rock-salt layer in each blocking layer. The Ae$_{n+2}$M$_{n}$O$_y$Fe$_2$As$_2$ and Ae$_{n+2}$M$_{n}$O$_y$Fe$_2$As$_2$ can be SC with T$_c$ ranging from 17 to 47 K \cite{cxh35,cxh36,cxh37}.
There are some other FeAs-based superconductors with quite complicated structures, such as Ca-Fe-Pt-As system Ca$_{10}$(Pt$_3$As$_8$)(Fe$_2$As$_2$)$_5$ (so called 10-3-8), Ca$_{10}$(Pt$_4$As$_8$)(Fe$_2$As$_2$)$_5$ (so called 10-4-8) \cite{cxh38,cxh39} and Ba$_2$Ti$_2$Fe$_2$As$_4$O \cite{cxh40} and so on.

\begin{table}[t]
        \centering
         \caption {Maximum temperatures of the SC transition under ambient pressure and lattice parameters of undoped compounds for some typical iron-pnictides.}
        \begin{tabular}{c|l|l|c|l|l} \hline
        Compound                  & Maximum T$_c$(K)                           & Space group          &
        a (\AA)                  & c (\AA)                           & Ref.          \\ \hline
        LiFeAs &  18  &   P4/nmm  & 3.775  &  6.353  &  [4]          \\        BaFe$_2$As$_2$ &38  &  I4/mmm & 3.963  & 13.017 & [13]                                         \\
        LaOFeAs& 41   &   P4/nmm & 4.035   &8.740 &  [24]                                         \\
       CeOFeAs & 41   &   P4/nmm  &3.996  & 8.648 &  [24]                                          \\
       PrOFeAs & 52   &   P4/nmm & 3.926  & 8.595 &  [24]                                           \\
        NdOFeAs &51.9 &   P4/nmm & 3.940 &  8.496 &  [24]                                             \\
       SmOFeAs & 55   &   P4/nmm & 3.940 &  8.496  & [24]                                            \\
        GdOFeAs& 53.5 &   P4/nmm & 3.915  & 8.435 &  [24]                                            \\
        TbOFeAs &48.5   & P4/nmm & 3.898  & 8.404  & [24]        \\ \hline
        \end{tabular}

        \label{tab:param}
\end{table}

\begin{figure*}[t] \includegraphics[scale=0.8]{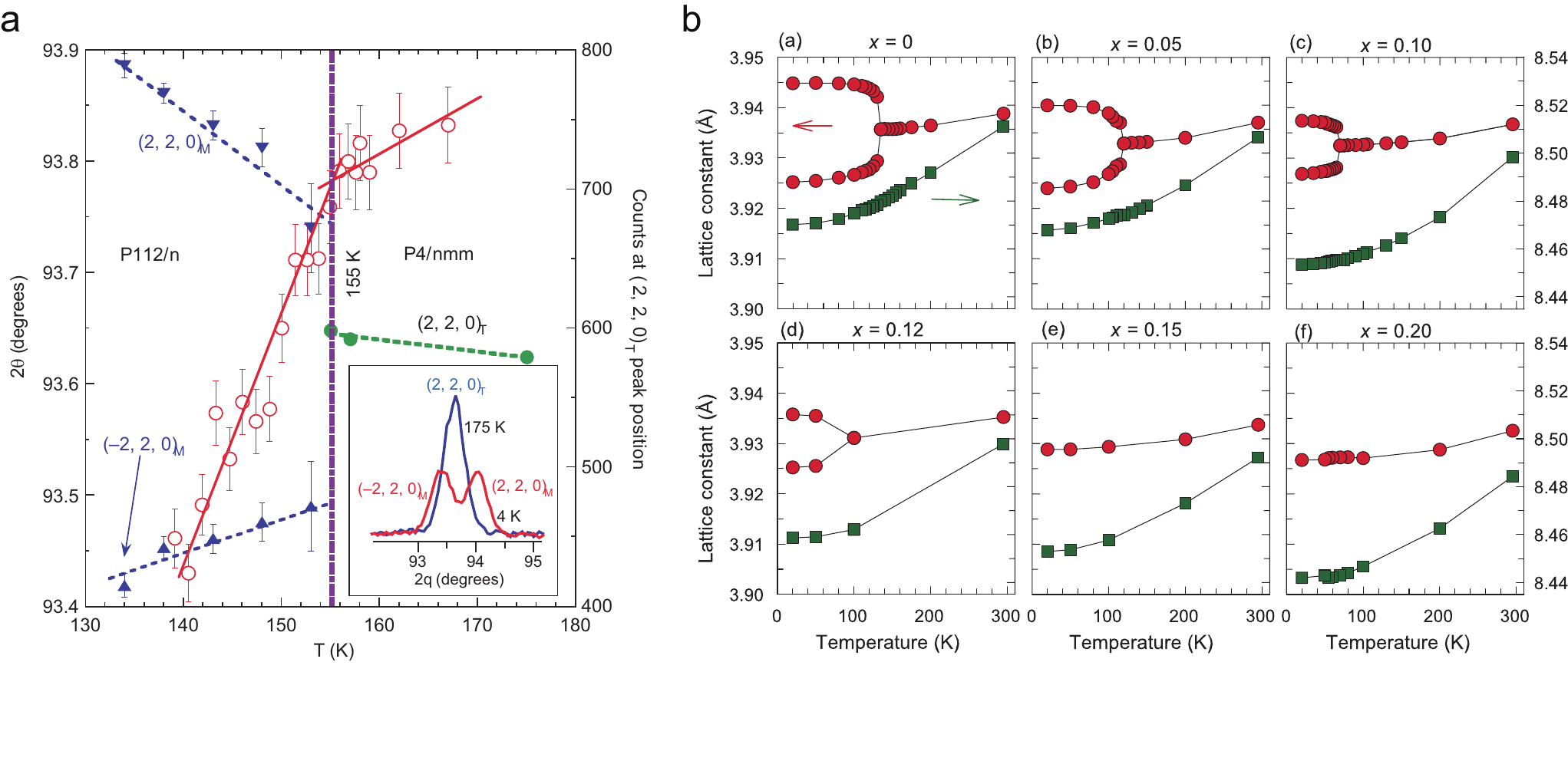}
\caption{
(a) Neutron scattering result on structural transitions on LaOFeAs \cite{cxh41}. (b) The temperature dependence of lattice constants for different F doping levels in SmFeAsO$_{1-x}$F$_x$ system.}
\label{fig:cxh2}
\end{figure*}

\begin{figure}[t] \includegraphics[scale=.2]{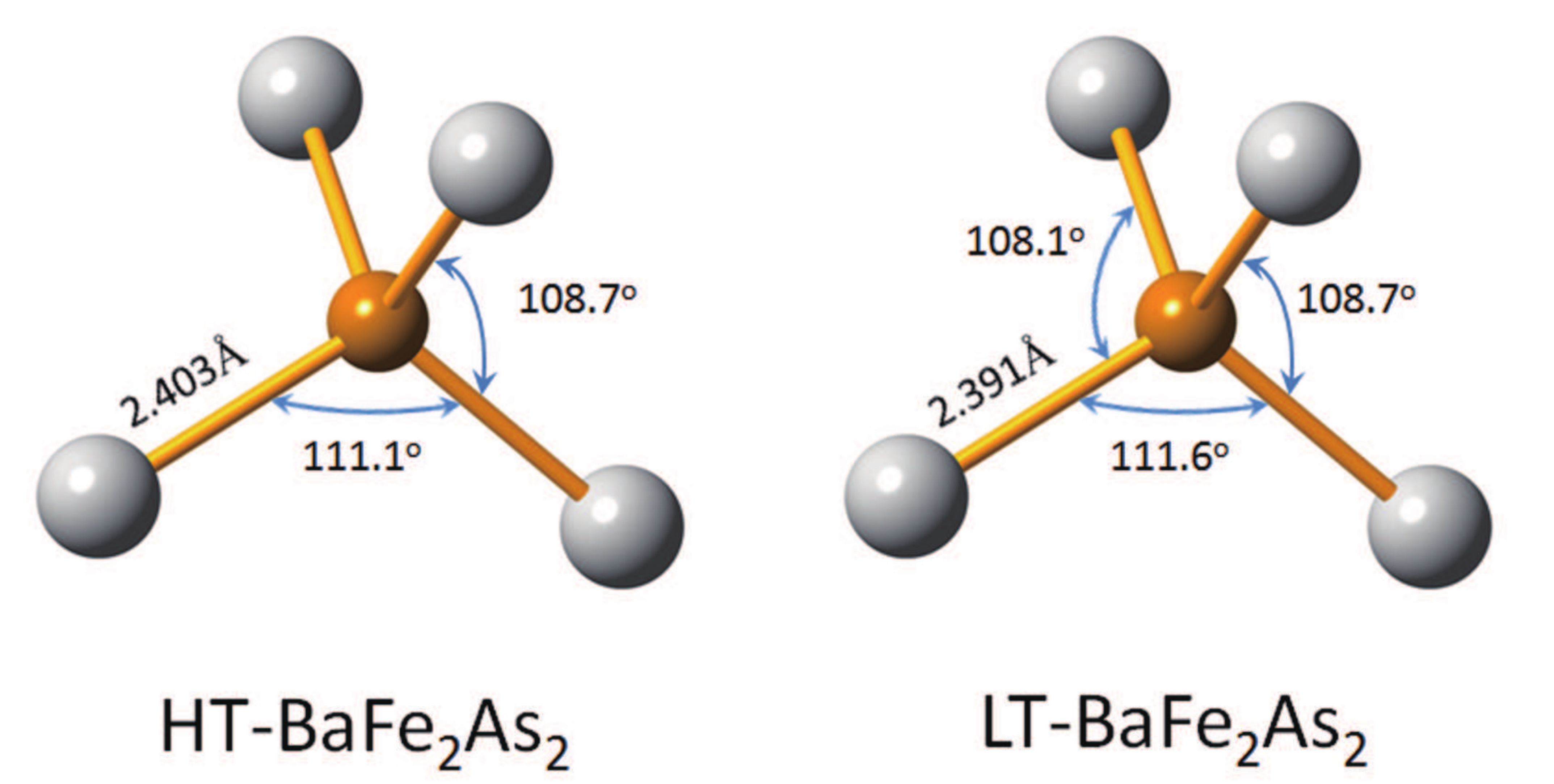}
\caption{
As-Fe-As bond angles of BaFe$_{2}$As$_{2}$ at high and low temperature respectively (data from Ref. \cite{122}).}
\label{fig:cxh3}
\end{figure}

There is a resistivity anomaly in the parent compound of LaFeAsO at around 150 K, which disappears as superconductivity emerges \cite{Hosono}. It was clarified later by de la Curz et al. through neutron scattering experiment that such an anomaly around 155 K could be attributed to the structural phase transition \cite{cxh41}. As shown in Fig. \ref{fig:cxh2}, a structural transition occurs around 155 K in undoped LaOFeAs. The space group of the low-temperature structure was  clarified to be  the orthorhombic $Cmma$ \cite{cxh42,cxh43}. The space group changes from the high-temperature tetragonal $P$4/$nmm$ to low-temperature orthorhombic $Cmma$, corresponding to a transformation from 5.70307 $\AA$$\times$5.70307 $\AA$ square network (for comparison, here we use \emph{a}$\sqrt{2}$, so that space group becomes $F$4/$mmm$) to 5.68262 $\AA$$\times$5.571043$\AA$ with a slight shrink of the c-axis lattice constant, as shown in the left panel of Fig. \ref{fig:cxh2} \cite{cxh44}. In the structural transition, chemical formulae in each unit cell change from 2 to 4 with a symmetry degradation. With decreasing temperature, the parent and slightly doped AeFe$_2$As$_2$ (Ae = Ca, Sr, Ba, and Eu) also undergo a structural transition from high-temperature tetragonal phase to low-temperature orthorhombic phase. The low-temperature orthorhombic phase has the space group of Fmmm \cite{122,cxh45}. Fig. \ref{fig:cxh3} shows that FeAs$_4$ tetrahedron distorts in the structural transition in BaFe$_2$As$_2$. The As-Fe-As angles around 108.7$^\circ$ become nonequivalent and evolve to two values of 108.1$^\circ$ and 108.7$^\circ$ respectively \cite{122}. Such a structural transition from high-temperature tetragonal symmetry to low-temperature orthorhombic symmetry occurs among the undoped and underdoped FeAs-based 111, 122 and 1111 phases.
\begin{figure*}[t]
\includegraphics[width=18cm]{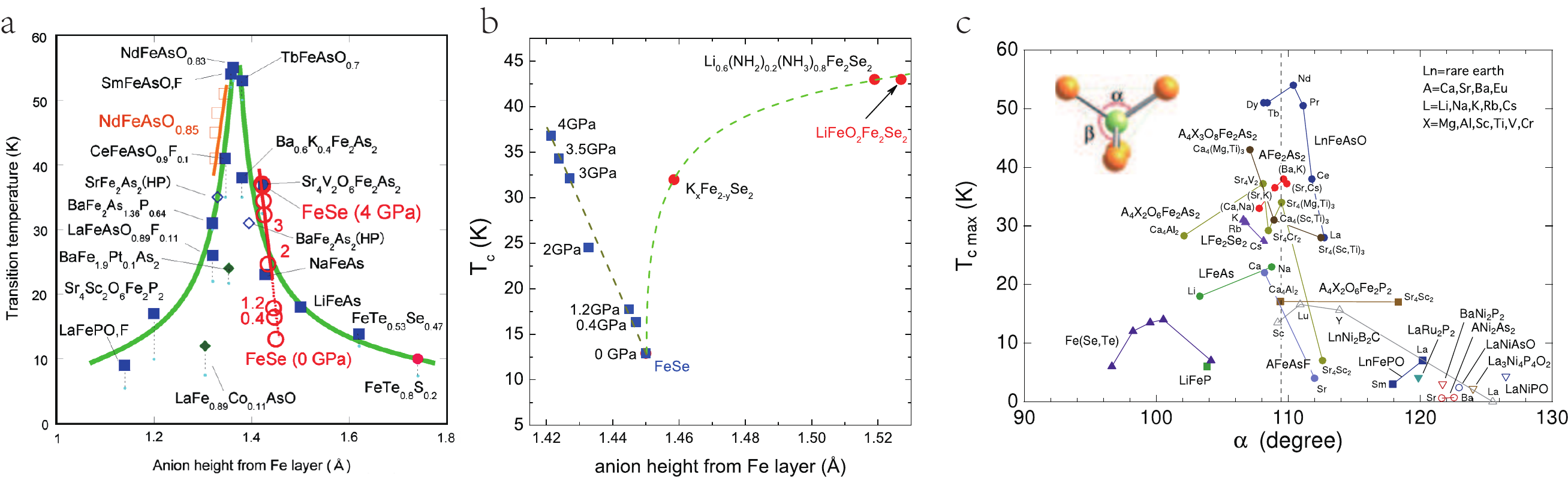}
\caption{
(a) Anion height dependence of $T_c$ in iron-based superconductors \cite{cxh49}. (b) Anion height dependence of $T_c$ in FeSe-derived superconductors.\cite{cxh32} (c) The relation between the As(top)-Fe-As(top) angle a and $T_c$ of iron-based superconductors \cite{cxh48}.}
\label{fig:cxh4}
\end{figure*}

\subsection{Relation between crystal structure and superconductivity}

By summarizing plenty of data about crystal structure and $T_{c}$ for iron-based superconductors, it is found that $T_{c}$ is related to structure parameters \cite{cxh48,cxh49,cxh4901}.  In particular, there is a close relation between the anion (As, P, Se, and Te) height from the Fe layer ($h$) and $T_{c}$, as shown in Fig. \ref{fig:cxh4}a \cite{cxh49}. $h$ depends on the type of anion, increasing in turn from FeP, FeAs, FeSe to FeTe. Due to the relative small $h$ in FeP based superconductors, their $T_{c}$s are usually lower than those in FeAs-based superconductors. For example, in La-1111 phase, as P is substituted by As, $T_{c}$ is enhanced dramatically from 7 to 26 K, due to the increase of $h$. For FeAs-based 1111 phase, as the substitutions of La by Nd and Sm increases  $h$ to around 1.38 \AA, $T_{c}$ increases dramatically from 26 to 56 K.  After crossing this maximum, the $T_{c}$ of TbFeAsO$_{0.7}$, Ba$_{0.6}$K$_{0.4}$Fe$_{2}$As$_{2}$, NaFeAs or LiFeAs decreases with  increasing $h$. The data of  optimally doped FeSe$_{1-x}$Te$_{x}$, FeSe$_{0.57}$Te$_{0.43}$, seem to also follow the same curve. As a result, such a dependence of $T_{c}$ on $h$ seems to be universal for 1111, 122, 111 and 11 iron-based superconductors. Though the maximum T$_{c}$ of the superconductors with thick blocking layer remains unconfirmed, the data of the 42622 superconductor obey the same universal curve, except for a small deviation, which may suggest that the enhancement of 2D character could induce a $T_{c}$ higher than 56 K. Note that for such a universal correspondance between $h$ and $T_{c}$, there is  an exception in FeSe-derived superconductors. As shown in Fig. \ref{fig:cxh4}b, for the FeSe-derived materials, a minimum of $T_{c}$ can be observed at $h \approx 1.45$ \AA \cite{cxh32}, instead of a maximum as shown in Fig. \ref{fig:cxh4}a. This may suggest some new underlying physics in FeSe-derived superconductors compared  to FeAs-based ones. The bond angle of As-Fe-As, which reflects the distortion of the FeAs$_{4}$ tetrahedron, was also thought to be closely related to superconductivity \cite{cxh50}. As shown in the Fig. \ref{fig:cxh4}c, the maximum $T_{c}$ was achieved when the FeAs$_{4}$ tetrahedron is in  perfectly regular, with the bond angle of 109.47$^\circ$.

\newpage

\section{The interplay between magnetism and superconductivity}
\begin{figure}[t] \includegraphics[scale=.45]{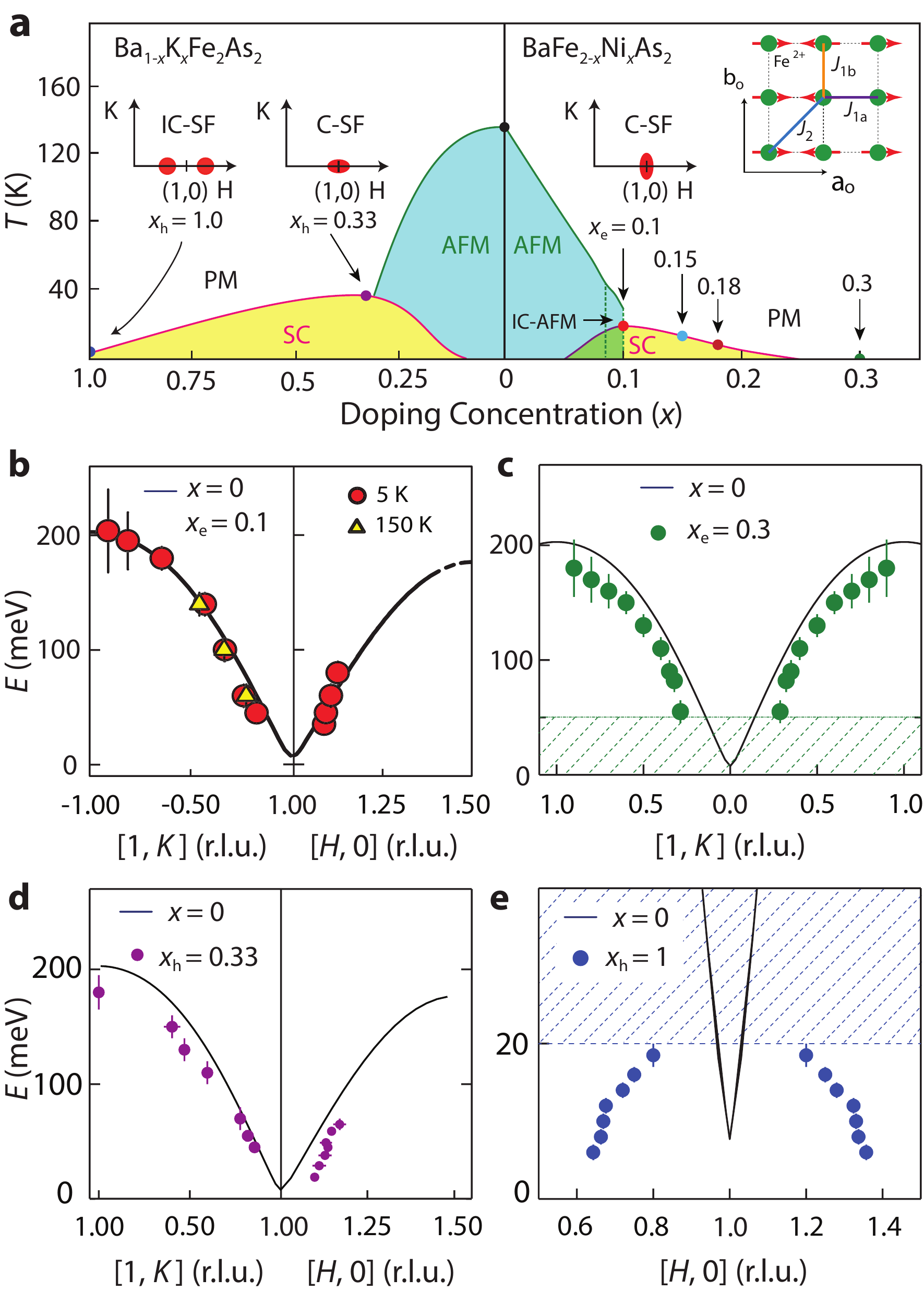}
\caption{
(a) The electronic phase diagram of electron and hole-doped BaFe$_2$As$_2$, where the arrows indicate the doping levels of inelastic neutron scattering
experiments. The right
inset shows crystal and AF spin structures of BaFe$_2$As$_2$ with marked the nearest ($J_{1a}$, $J_{1b}$) and next nearest
neighbor ($J_2$) magnetic exchange couplings. The inset above $x_e=0.1$ shows the transversely elongated ellipse representing the low-energy
spin excitations
in electron-doped BaFe$_{2-x}$Ni$_x$As$_2$ in the $(H,K)$ plane of the reciprocal space.
The left insets show the evolution of low-energy spin excitations in hole-doped Ba$_{1-x}$K$_x$Fe$_2$As$_2$ in the $(H,K)$ plane.
C-SF and IC-SF indicate commensurate and incommensurate spin fluctuations, respectively.
(b-e) The solid lines in the figure are spin wave dispersions of the undoped BaFe$_2$As$_2$ along the two high-symmetry directions.
The symbols in (b), (c), (d), and (e) are dispersions of spin excitations for BaFe$_{1.9}$Ni$_{0.1}$As$_2$, BaFe$_{1.7}$Ni$_{0.3}$As$_2$,
Ba$_{0.67}$K$_{0.33}$Fe$_2$As$_2$, and KFe$_2$As$_2$, respectively \cite{MWang13}.
The shaded areas indicate vanishing spin excitations.
}
\label{fig:dai1}
\end{figure}

Soon after the discovery of iron pnictide superconductor LaFeAsO$_{1-x}$F$_x$ with $T_c=26$ K \cite{Hosono},
band calculations
based on transport and optical conductivity measurements
predicted the presence of a collinear AF [or spin-density-wave (SDW)]  state in the parent compounds \cite{jdong}, which was subsequently confirmed by neutron scattering experiments as shown in the inset of Fig. \ref{fig:dai1} \cite{cxh41}.  The same collinear AF state was later found in
the parent compounds of most iron pnictide superconductors \cite{cxh45}.
Although the electronic phase diagrams for different families of iron-based superconductors can
 be somewhat different \cite{Johnston}, they all share the common feature of an AF ordered parent compound \cite{uemura}.
This has inspired many  to believe that the magnetic excitations play an important role in the mechanism of the high-$T_c$ superconductivity in iron-based supercondutors \cite{scalapino}.
To test if this is indeed the case, systematic investigation on the magnetic order and spin excitations through out the
phase diagram of different families of iron-based superconductors is essential.
 Figure \ref{fig:dai1} shows the electronic phase diagram of electron and hole-doped BaFe$_2$As$_2$ iron pnictides determined from transport and neutron scattering experiments \cite{AIGoldman4,MYWang1,MYWang2,AIGoldman5,RJMcQueeney1,HQLuo3,XYLu,savci,Dai2012}.  For electron-doped BaFe$_{2-x}$Ni$_x$As$_2$, the maximum $T_c=20$ K is around $x\approx 0.1$ and
 superconductivity ceases to exist for $x\geq0.25$ \cite{lli}.  For hole-doped Ba$_{1-x}$K$_x$Fe$_2$As$_2$, superconductivity exists in the entire phase diagram with maximum $T_c=38$ K near
 $x\approx 0.33$ and $T_c=2$ K for pure KFe$_2$As$_2$ \cite{rotter}.  The arrows in the figure indicate iron pnictides where
 spin excitations in the entire Brillouin zone have been mapped out by
 inelastic neutron scattering (INS) experiments \cite{LWHarriger,MSLiu12,HQLuo6,MWang13,CLZhang11,JPCastellan,CHLee11}.

\subsection{Effect of electron-doping to spin waves of iron pnictides}
We first discuss the electron-doping evolution of spin excitations in BaFe$_{2-x}$Ni$_x$As$_2$.  With the development of neutron time-of-flight spectroscopy, the entire spin wave spectra
were obtained in CaFe$_2$As$_2$ \cite{jzhao09} and BaFe$_2$As$_2$ \cite{LWHarriger} soon after the availability of single crystals of these materials. The solid lines in Figure \ref{fig:dai1}b show the dispersion of spin waves in BaFe$_2$As$_2$ along the $[1,K]$ and $[H,0]$ directions in reciprocal space, where
the collinear AF order occurs at the $Q_{AF}=(1,0)$  wave vector position \cite{LWHarriger}.  Upon electron-doping to induce optimal superconductivity, spin excitations become broader at low-energies ($E\le 80$ meV) while remain unchanged at high energies ($E>80$ meV) \cite{MSLiu12}.  The low-energy spin excitations couple to superconductivity via a collective spin excitation mode termed neutron spin resonance \cite{ADChristianson,MDLumsden09,SXChi,SLLi09}, seen also in copper oxide superconductors \cite{eschrig}. The red circle and yellow upper triangle
symbols in Fig. \ref{fig:dai1}b show spin excitation dispersions of the optimally electron doped BaFe$_{1.9}$Ni$_{0.1}$As$_2$ at
$T=5$ K and 150 K, respectively \cite{MSLiu12}.  With further electron-doping, superconductivity is suppressed for $x\geq0.25$ \cite{lli},
Figure \ref{fig:dai1}c shows the dispersions of spin excitations of
BaFe$_{1.7}$Ni$_{0.3}$As$_2$ compared with that of the undoped BaFe$_2$As$_2$ \cite{LWHarriger,MWang13}.  A large spin gap forms for energies below $\sim$50 meV as shown in the
dashed line region. The dispersions of spin excitations are also softer than that of BaFe$_2$As$_2$ \cite{LWHarriger,MWang13}.
Figures \ref{fig:dai1}d and \ref{fig:dai1}e show the dispersions of spin excitations
for optimally hole-doped Ba$_{0.67}$K$_{0.33}$Fe$_2$As$_2$ and hole over-doped KFe$_2$As$_2$, respectively.
The solid lines are the spin wave dispersions of BaFe$_2$As$_2$ \cite{LWHarriger,MWang13}.

\begin{figure*}[t!] \includegraphics[scale=.33]{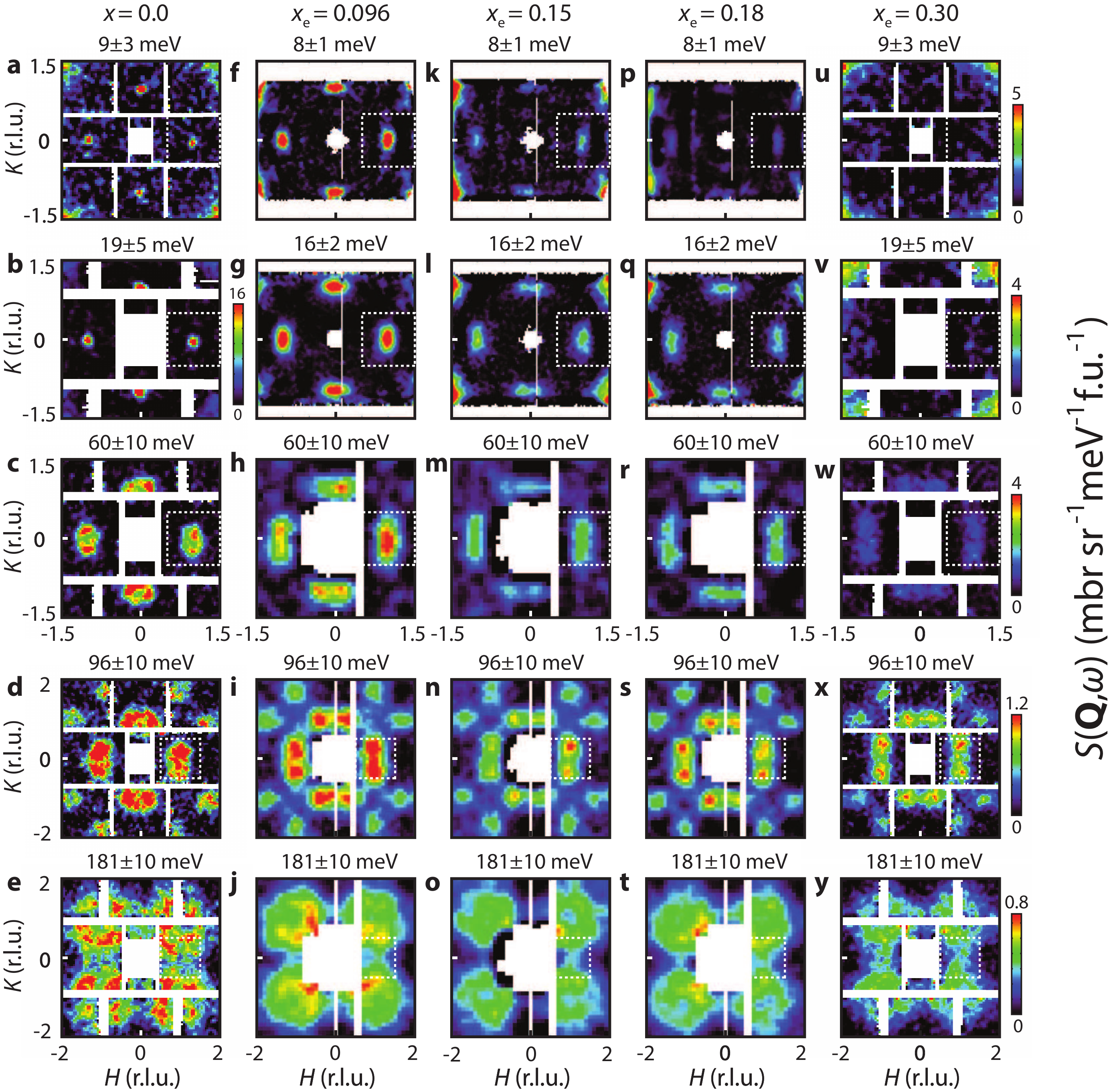}
\caption{
Constant-energy slices through magnetic excitations of electron-doped iron pnictides at different energies.
The color bars represent the vanadium
normalized absolute spin excitation intensity in the units of mbarn sr$^{-1}$meV$^{-1}$f.u.${-1}$.
(a-e) Spin waves of BaFe$_2$As$_2$ at excitation energies of $E=9\pm 3$, $19\pm 5$, $60\pm10$, $96\pm 10$, and $180\pm 10$ meV \cite{LWHarriger}.  Spin waves peak at
the AF ordering wave vectors $Q_{AF}=(\pm 1,0)$ in the orthorhombic notation.  Spin waves are also seen at $Q_{AF}\approx (0,\pm 1)$ due to the twin domains of the orthorhombic structure.
(f-j) Two-dimensional images of spin excitations for BaFe$_{1.904}$Ni$_{0.096}$As$_2$ at $E=8\pm 1$, $16\pm 2$, $60\pm 10$, $96\pm 10$,
$181\pm 10$ meV.  Identical slices as that of (f-j) for (k-o) BaFe$_{1.85}$Ni$_{0.15}$As$_2$ and (p-t) BaFe$_{1.82}$Ni$_{0.18}$As$_2$ \cite{HQLuo3}.
(u-y) Constant-energy slices through magnetic excitations of electron overdoped doped nonsuperconducting
 BaFe$_{1.7}$Ni$_{0.3}$As$_2$ at $E=9\pm 3$, $19\pm 5$, $60\pm 10$, $96\pm 10$,
$181\pm 10$ meV \cite{MWang13}.
}
\label{fig:dai2}
\end{figure*}

Figure \ref{fig:dai2} summarizes the evolution of the two-dimensional constant-energy images of spin excitations in the $(H, K)$ plane at different energies
 as a function of electron-doping from the undoped AF parent compound
BaFe$_2$As$_2$ to overdoped nonsuperconducting BaFe$_{1.7}$Ni$_{0.3}$As$_2$ \cite{LWHarriger,HQLuo3,MWang13}.
In the undoped case, there is an anisotropy spin gap
below $\sim$15 meV so there are essentially no signal at $E=9\pm 3$ meV (Fig. \ref{fig:dai2}a). Upon electron-doping to suppress static AF order and induce near optimal
superconductivity with $x=x_e=0.096$, the spin gap is suppressed and low-energy spin excitations are dominated by a resonance that couples with
superconductivity (Fig. \ref{fig:dai2}f) \cite{MDLumsden09,SXChi,SLLi09,HQLuo3}.  Moving to electron overdoped side with reduced superconductivity
in BaFe$_{2-x}$Ni$_x$As$_2$ with $x=0.15$ ($T_c=14$ K) and 0.18 ($T_c=8$ K), spin excitations at $E=8\pm 1$ meV become weaker and more transversely elongated (Figs. \ref{fig:dai2}k and \ref{fig:dai2}p) \cite{HQLuo3}.
Finally on increasing electron doping level to $x=0.3$ with no superconductivity, a large spin gap forms in the low-energy excitations spectra (Fig. \ref{fig:dai2}u).  Figures \ref{fig:dai2}b-\ref{fig:dai2}e,
\ref{fig:dai2}g-\ref{fig:dai2}j, \ref{fig:dai2}i-\ref{fig:dai2}o, \ref{fig:dai2}q-\ref{fig:dai2}t,\ref{fig:dai2}v-y show spin excitations at different energies for BaFe$_{2-x}$Ni$_x$As$_2$ with $x=0,0.096,0.15,0.18,$ and 0.30, respectively.  While spin excitations at
energies below $E=96\pm 10$ meV change rather dramatically with increasing electron-doping, high energy spin excitations remain similar and only soften slightly.

\subsection{The effect of hole-doping to the spin excitations of iron pnictides}
\begin{figure*}[t] \includegraphics[width=\columnwidth,trim= 60 245 0 0, clip=true]{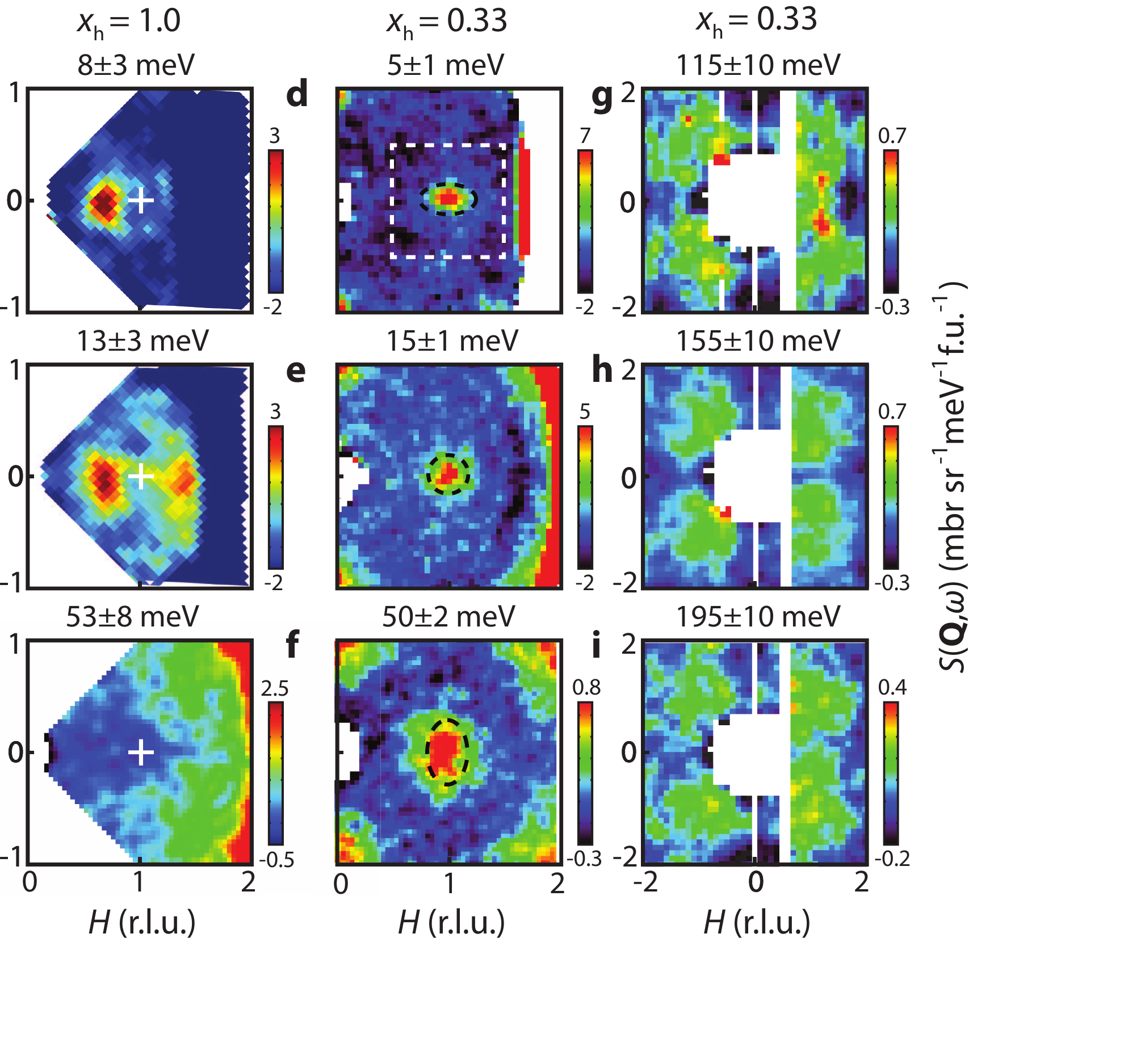}
\caption{
Two-dimensional images of spin excitations at different energies for hole-doped KFe$_2$As$_2$ at 5 K.
(a) $E=8\pm 3$ meV obtained with $E_i=20$ meV along the $c$-axis.
The right side incommensurate peak is obscured by background scattering.
 (b) $13\pm3$ with $E_i=35$ meV,  and (c) $53\pm 8$ meV with $E_i=80$ meV.
For  Ba$_{0.67}$K$_{0.33}$Fe$_2$As$_2$ at $T=45$ K, images of spin excitations at (d) $E=5\pm 1$ meV obtained with $E_i=20$ meV, (e) $15\pm 1$ meV with
$E_i=35$ meV, and (f) $50\pm 2$ meV obtained with $E_i=80$ meV.
Spin excitations
of Ba$_{0.67}$K$_{0.33}$Fe$_2$As$_2$ at energy transfers
(g) $115\pm 10$ meV; (h) $155\pm 10$ meV (i) $195\pm 10$ meV obtained
with $E_i=450$ meV, all at 9 K. Wave vector dependent backgrounds have been subtracted from the images \cite{MWang13}.
}
\label{fig:dai3}
\end{figure*}

Figure \ref{fig:dai3} shows the evolution of spin excitations in the similar two-dimensional constant-energy images as a function of hole-doping.
For pure KFe$_2$As$_2$, incommensurate spin excitations along the longitudinal direction are seen at $E=8\pm 3$ (Fig. \ref{fig:dai3}a) and
$13\pm 3$ (Fig. \ref{fig:dai3}b) \cite{CHLee11}.  Upon further increasing energy, no clear magnetic scattering can be seen (Fig. \ref{fig:dai3}c).
For optimally hole-doped Ba$_{0.67}$K$_{0.33}$Fe$_2$As$_2$, the low-energy spin excitations change from transversely elongated ellipses as shown in Fig. \ref{fig:dai2}f to
longitudinally elongated ellipse at $E=5\pm1$ meV (Fig. \ref{fig:dai3}d).  On increasing the energy to the neutron spin resonance energy of $E=15\pm 1$ meV, spin
excitations become isotropic in reciprocal space (Fig. \ref{fig:dai3}e).  Spin excitations become transversely elongated again for energies above $E=50\pm 2$ meV (Figs. \ref{fig:dai3}f-\ref{fig:dai3}i), very similar to
spin excitations in electron-doped BaFe$_{2-x}$Ni$_x$As$_2$ (Fig. \ref{fig:dai2}).  From data presented in Figs. \ref{fig:dai2} and \ref{fig:dai3}, one can establish the basic trend in the evolution of spin excitations via electron
and hole doping to the parent compound BaFe$_2$As$_2$.  While electron-doping appears to
 mostly modify the low-energy spin excitations and make them more transversely elongated with increasing electron counts, high-energy spin excitations do not change dramatically.  Therefore,
 the Fermi surface modifications due to electron doping affect mostly the low-energy spin excitations, suggesting that they are arising from itinerant electrons.  The high-energy spin excitations weakly dependent on electron-doping induced Fermi surface changes
 are most likely arising from localized moment.  The lineshape change from transversely to longitudinally
elongated ellipse in low-energy spin excitations of iron pnictides upon hole-doping is consistent with the random phase approximation calculation
of the doping dependence of the nested hole and electron Fermi surfaces \cite{CLZhang11,jtpark10}.  The absence of dramatic changes in high-energy spin excitations again suggests
the presence of local moments independent of Fermi surface changes induced by electron or hole-doping. Comparing with resonant inelastic X-ray scattering (RIXS) results on hole-doped
Ba$_{0.6}$K$_{0.4}$Fe$_2$As$_2$ \cite{kjzhou}, we note that dispersion determined from RIXS is consistent with neutron scattering while the intensity is lower. At present, it is unclear how to understand the intensity of the RIXS measurements.

\subsection{Evolution of local dynamic susceptibility as a function of electron and hole-doping}
\begin{figure}[h!] \includegraphics[scale=0.45]{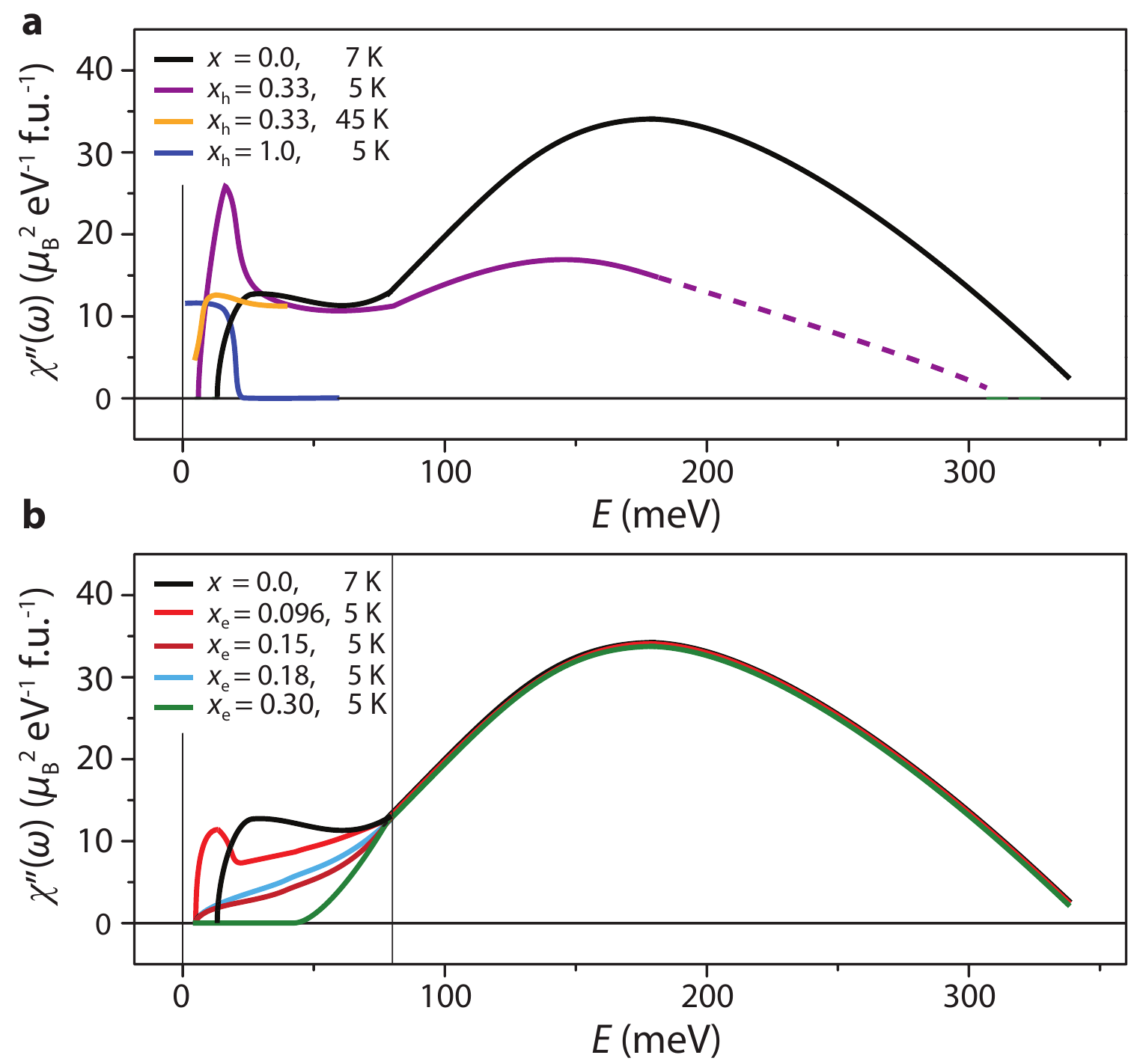}
\caption{
Energy and temperature dependence of the local dynamic susceptibility  $\chi^{\prime\prime}(\omega)$ for (a)
BaFe$_2$As$_2$, Ba$_{0.67}$K$_{0.33}$Fe$_2$As$_2$, KFe$_2$As$_2$, and (b) BaFe$_{2-x_e}$Ni$_{x_e}$As$_2$ with $x_e=0,0.096,0.15,0.18,0.3$.
The intensity is in absolute unit of $\mu_B^2$eV$^{-1}$f.u.$^{-1}$ obtained by integrating the $\chi^{\prime\prime}(Q,\omega)$ in the
dashed regions specified in Figs. \ref{fig:dai2} and \ref{fig:dai3}.
}
\label{fig:dai4}
\end{figure}

To quantitatively determine the electron and hole-doping evolution of the spin excitations in iron pnictides, one can estimate the energy dependence of the
local dynamic susceptibility per formula unit, defined as $\chi^{\prime\prime}(\omega)=\int{\chi^{\prime\prime}({\bf
    q},\omega)d{\bf q}}/\int{d{\bf q}}$,
where $\chi^{\prime\prime}({\bf q},\omega)=(1/3) tr( \chi_{\alpha \beta}^{\prime\prime}({\bf q},\omega))$ \cite{MSLiu12,lester10}.  The dashed squares in Figs. \ref{fig:dai2} and \ref{fig:dai3} show the
integration region of the
local dynamic susceptibility in reciprocal space. Figure \ref{fig:dai4}a and \ref{fig:dai4}b summarizes the energy dependence of the
local dynamic susceptibility for hole (Ba$_{1-x_h}$K$_{x_h}$Fe$_2$As$_2$ with
$x_h=0,0.33,1$) and electron (BaFe$_{2-x_e}$Ni$_{x_e}$As$_2$ with $x_e=0,0.096,0.15,0.18,0.3$) doped iron pnictides.
From Figure \ref{fig:dai4}a, we see that the effect of hole-doping near optimal superconductivity is to
suppress high-energy spin excitations and transfer spectral weight to lower-energies.  The intensity changes across $T_c$ for hole-doped Ba$_{0.67}$K$_{0.33}$Fe$_2$As$_2$ are much larger than that of the electron-doped BaFe$_{1.9}$Ni$_{0.1}$As$_2$ \cite{MSLiu12}.  As a function of increasing electron-doping, the local dynamic susceptibility at low-energies decreases and finally vanishes for electron-overdoped nonsuperconducting  BaFe$_{1.7}$Ni$_{0.3}$As$_2$ with no hole-like Fermi surface \cite{HQLuo3,MWang13}.  This again confirms the notion that superconductivity in iron pnictide is associated with itinerant electron and low-energy spin excitation coupling  between the nested hole and electron Fermi surfaces \cite{MWang13}.

\subsection{Correlation between spin excitations and superconductivity}
In conventional BCS superconductors \cite{bcs}, superconductivity occurs via electron-lattice coupling below $T_\mathrm{c}$.
The SC condensation energy $E_\mathrm{c}$ ($=-N(0)\Delta^2/2$ and $\Delta\approx 2\hbar\omega_\mathrm{D} e^{-1/N(0)V_0}$, where $N(0)$ is the electron density of states at zero temperature) and $T_\mathrm{c}$ are controlled by the strength of
the Debye energy $\hbar\omega_\mathrm{D}$ and electron-lattice coupling $V_0$ \cite{bcs,chester,schrieffer}.
For unconventional superconductors derived from electron and hole-doping to their AF ordered parent compounds,
short-range spin excitations may mediate electron pairing for superconductivity \cite{scalapino}.  Here
the SC condensation energy should be accounted for by the change in magnetic exchange energy between the normal (N) and superconducting (S)
phases at zero temperature via $\Delta E_\mathrm{ex}(T)=2J[\left\langle {\bf S}_{i+x}\cdot{\bf S}_{i}\right\rangle_\mathrm{N}-
\left\langle {\bf S}_{i+x}\cdot{\bf S}_{i}\right\rangle_\mathrm{S}]$, where $J$ is the nearest neighbor magnetic exchange coupling,
$\left\langle {\bf S}_{i+x}\cdot{\bf S}_{i}\right\rangle$ is the dynamic spin
susceptibility in absolute units at temperature $T$, and $S({\bf Q},E=\hbar\omega)$ is related to the imaginary part of the dynamic
susceptibility $\chi^{\prime\prime}({\bf Q},\omega)$
via $S({\bf Q},\omega)=[1+n(\omega,T)]\chi^{\prime\prime}({\bf Q},\omega)$ with $[1+n(\omega,T)]$ being the Bose population factor \cite{scalapino}.

Since the dominant magnetic exchange couplings are isotropic nearest neighbor exchanges
for copper oxide superconductors \cite{tranquada,headings}, the magnetic exchange energy $\Delta E_\mathrm{ex}(T)$ can be directly estimated
using the formula through carefully measuring of $J$ and
the dynamic spin susceptibility in absolute units between the normal and SC states
\cite{demler,woo,dahm}.  For heavy Fermion \cite{stockert} and iron pnictide superconductors \cite{MWang13},
 one has to modify the formula to include both
the nearest neighbor and next nearest neighbor magnetic exchange couplings. The calculations of the magnetic exchange energies in CeCu$_2$Si$_2$ \cite{stockert} and
optimally hole-doped Ba$_{0.67}$K$_{0.33}$Fe$_2$As$_2$ \cite{MWang13} reveal that they are large enough to account for the SC condensation energy, thus suggesting
that spin excitations could be the driving force for mediating electron pairing for superconductivity. These results are consistent with NMR experiments \cite{flning}, where the presence low-energy spin excitations is found to be associated with Fermi surface nesting and the absence of nesting in electron overdoped iron pnictides supresses the low-energy spin excitations.  These results are also consistent with the absence of spin excitations in non-superconducting
collapsed tetragonal phase of CaFe$_2$As$_2$ from inelastic neutron scattering measurements \cite{jhsoh}.

\newpage

\section{Electronic properties}



Electronic properties of the iron based superconductor is critical for understanding its phase diagram, mechanism and transport behaviors.   In this section, we focus on  the electronic properties of these compounds obtained by
ARPES, in link with some results from scanning tunneling microscopy/spectroscopy (STM/STS) measurements.   We note that  more detailed reviews of the ARPES results can be found in Refs.~\onlinecite{AAPPS,Ye_CPB}.

The electronic structure of the iron based superconductors and related materials is characterized by the  multi-band and multi-orbital nature \cite{ZYLu}.  Based on their Fermi surface topology, one could generally divide them into two categories: 1. systems with both electron and hole Fermi surfaces, and 2. systems with only electron Fermi surfaces. Most of the iron pnictides and  Fe(Te,Se) bulk materials belong to the first category,\cite{lxyang,lxyangPRB,Hecheng,FeTePRB,ChenPRB} while A$_x$Fe$_{2-y}$Se$_2$, single layer FeSe on STO, and heavily electron doped iron pnictides belong to the second category \cite{XJZhou_KFeSe,KFeSe,tan,zirong}. Since most of the iron based superconductors are in the first category, we will discuss its electronic properties in the following first three subsections, while those of the second category will be discussed in the fourth subsection. In the last subsection, we will discuss the role of correlations and some  overall understandings of the electronic structure.

\subsection{Basic electronic structure}

\begin{figure}[t]
\includegraphics[width=8.7cm]{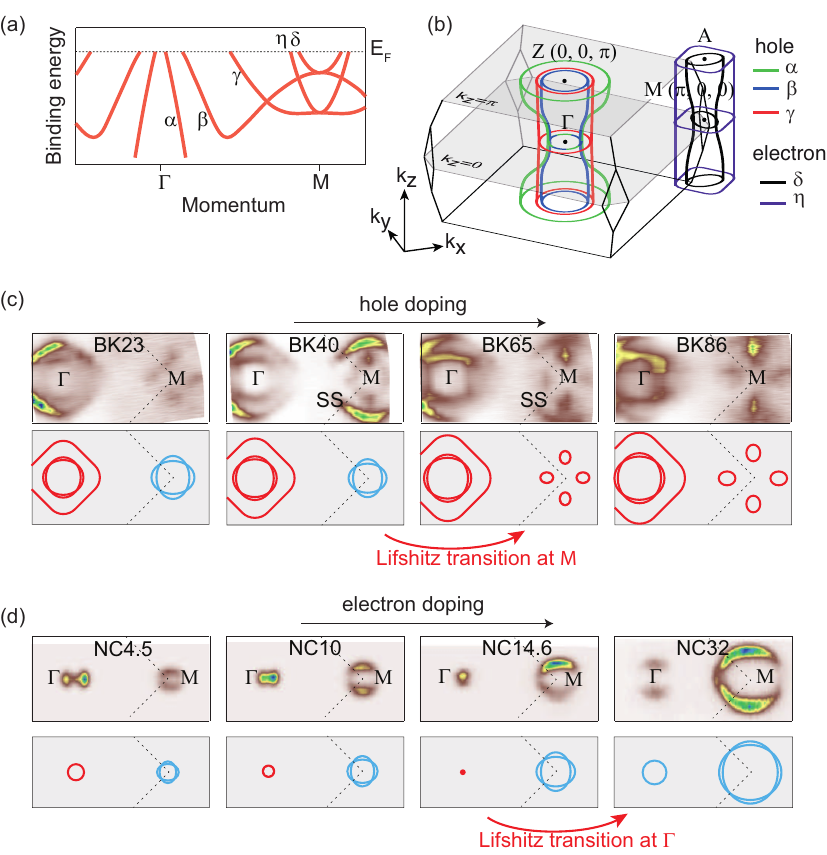}
\caption{(a) Cartoon of the band structure in iron-pnictides. (b) The typical three-dimensional Fermi surface of iron-pnictides BaFe$_2$(As$_{0.7}$P$_{0.3}$)$_2$ \cite{Zhangnode}. (c) The doping dependence of Fermi surface topology taken in
Ba$_{1-x}$K$_x$Fe$_2$As$_2$. The upper panels are the photoemission intensity distribution at $E_F$. The low panels are the obtained Fermi surface. SS is the abbreviation of surface state. The red and blue lines illustrate the hole pockets and electron pockets, respectively. (d) is the same as panel (c), but taken in NaFe$_{1-x}$Co$_x$As. } \label{basic}
\end{figure}

The low energy electronic structure of iron-based superconductors are dominated by the Fe 3$d$ states \cite{Singh}. The unit cell contains two Fe ions, because there are pnictogen and chalcogen ions above and below the iron plane. As a result, there would be totally ten $3d$ states and thus ten bands. However, it was found both by polarization-dependent ARPES experiments and density functional theory calculations that the Fermi surface is usually composed of three hole-like Fermi surfaces near the zone center and two electron Fermi surfaces around the zone corner. The $d_{xy}$, $d_{xz}$ and $d_{yz}$ orbitals are the main contributor to the states near the Fermi energy $E_F$ \cite{YZhangBaCo,IIMazin}. Fig.~\ref{basic}(a) illustrates a typical band structure  along the $\Gamma$-M direction, and Fig.~\ref{basic}(b) shows the measured Fermi surface of  BaFe$_2$(As$_{0.7}$P$_{0.3}$)$_2$ in its three-dimensional Brillouin zone \cite{zirongye}.
The Fermi surface topology and band structure are rather similar for the 11, 111, 122, 1111 series of iron-based superconductors \cite{lxyang,lxyangPRB,Hecheng,FeTePRB,ChenPRB}.

Strictly speaking, the iron-based superconductors are three dimensional materials, however, the electron hopping along the $c$ direction is not strong, so that the warping of the Fermi surface along the $k_z$ direction is not so obvious for most of the bands. However, the $\alpha$ band shows strong $k_z$  dependence, and its Fermi surface exhibits large warping in Fig.~\ref{basic}(b).

The carrier density of iron-based superconductors can be tuned by doping in the charge reservior layer, or at the Fe site. Electron and hole doping can be varied over a large range, which in turn varies the chemical potential and the Fermi surface. Fig.~\ref{basic}(c) shows the hole-doping evolution of the Fermi surface sheets in  Ba$_{1-x}$K$_x$Fe$_2$As$_2$, and the electron-doping evolution of the Fermi surface sheets in NaFe$_{1-x}$Co$_x$As.
With sufficient doping, Lifshitz transitions of the Fermi surface topology eventually occur in both cases \cite{HDing_KFeAs,zirong}.

In the case of  isovalent doping, such as Ru for Fe,  P for As, or Te for Se, the Fermi surface volume usually does not change, or change slightly for some unknown reason, but the individual Fermi surface sheets may change noticebly by the induced  chemical pressure \cite{CLiu,zirongye}. The band renormalization factor also generally decreases with doping, which indicates the weakened correlation.


Compared with the cuprate superconductors, which is a doped Mott insulator, the iron-based superconductors are  less sensitive to impurities. However, the large amount of dopants will cause serious scattering of the quasiparticles that needs to be taken into account in understanding the transport and SC properties. The $d_{xy}$-based $\gamma$ band near the zone center are somehow more sensitive to impurities \cite{zirong}. The impurity scattering also strongly depends  on the locations of the dopants, which in the increasing order is off-plane, at the pnictogen/chalcogen site, and at the iron site \cite{Uchida1,zirong}.


\begin{figure*}[]
\includegraphics[width=13.5cm]{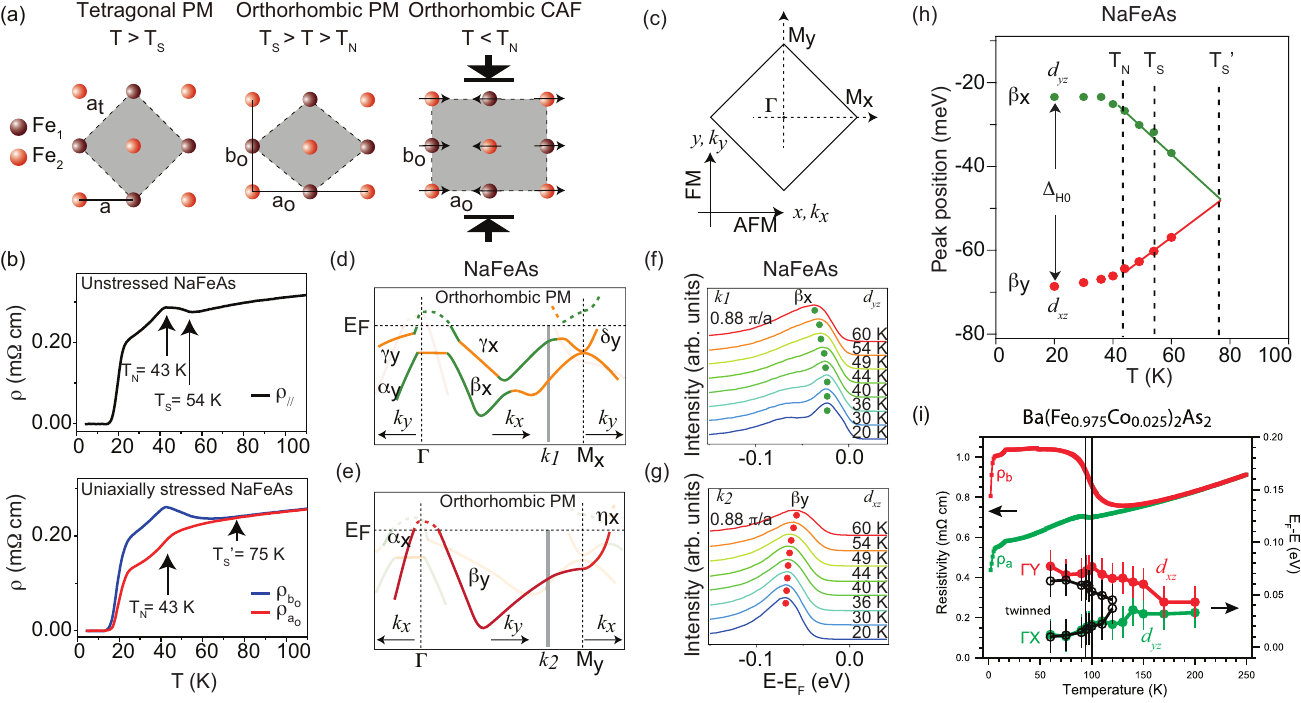}
\caption{(a) Cartoon of the lattice and spin structure in tetragonal paramagnetic (PM), orthorhombic PM, and orthorhombic CAF state for iron-pnictides. The large black arrows show the direction of the uniaxial pressure applied in the mechanical detwinning process. (b) The temperature dependent resistivity of unstressed and uniaxially stressed NaFeAs, respectively. (c) The definition of the projected two-dimensional Brillouin zone for NaFeAs. The $x$ and $y$ axes are defined along the iron-iron directions. (d) The band structure in the orthorhombic PM state along $\Gamma$-M$_x$ and some other high symmetry directions,  where only $d_{yz}$- and $d_{xy}$-dominated bands are highlighted.  (e) is the same as panel (d), but mainly  along the $\Gamma$-M$_y$ direction, with the $d_{xz}$-dominated bands highlighted. (f) and (g) The temperature dependence of the EDCs at $k_1$ and $k_2$,  as indicated by the gray line in panels (d) and (e), respectively. (h) The peak positions of the $\beta_x$ and $\beta_y$ bands as functions of temperature. The maximal observable separation between $\beta_x$ and $\beta_y$ at the same momentum value (\textit{i.e.} $|k_x|=|k_y|$) near $M_x$ and $M_y$ respectively is defined as $\Delta_{H}$.  $\Delta_{H}$ is a function of temperature, and its low temperature saturated value is defined as $\Delta_{H0}$. (i) Energy position of the $d_{xz}$ and $d_{yz}$ bands as a function of temperature, measured on both detwinned and unstressed Ba(Fe$_{0.975}$Co$_{0.025}$)$_2$As$_{2}$, compared with resistivity measurements.\cite{FisherResistivity} Data are taken from Ref.~\onlinecite{NaFeAsPRB} and  Ref.~\onlinecite{MYi}.} \label{nematic}
\end{figure*}

\begin{figure*}[]
\includegraphics[width=13.5cm]{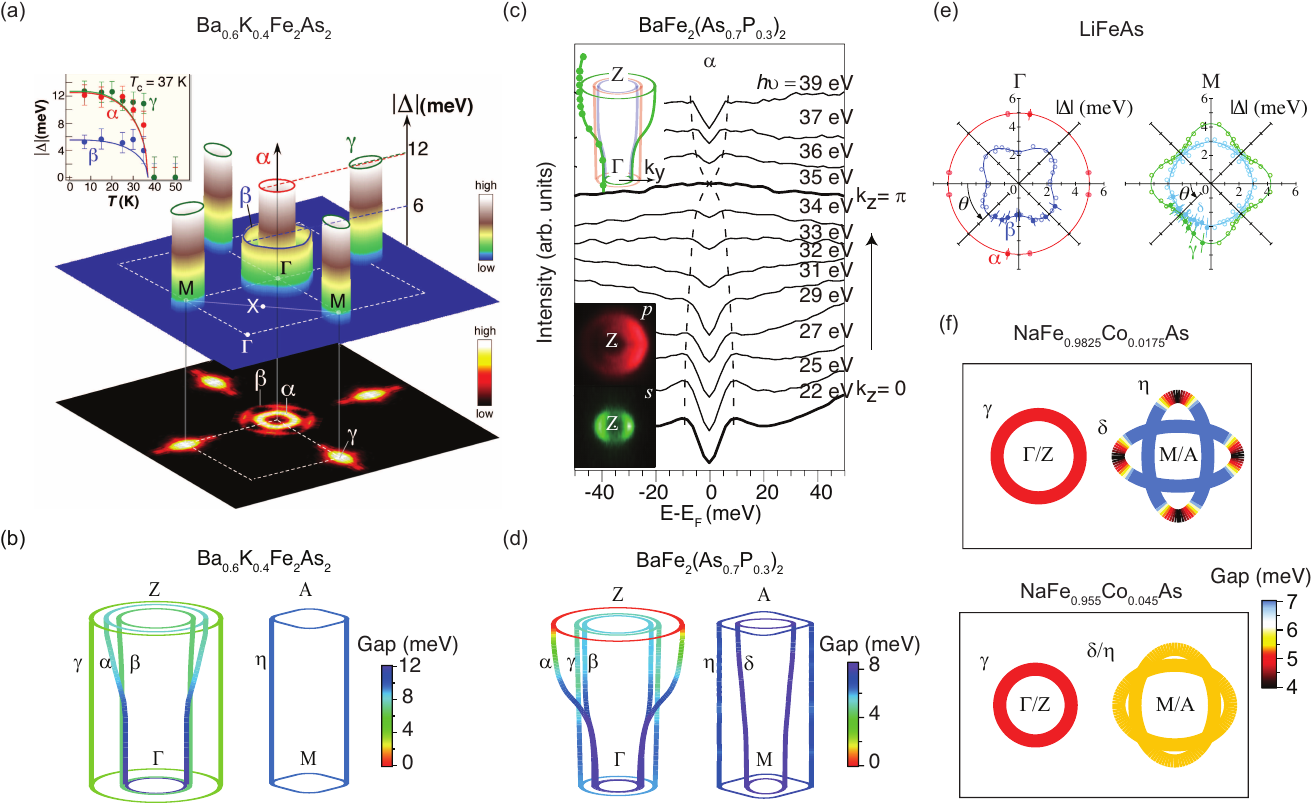}
\caption{(a) The illustration of the in-plane gap distribution on the Fermi surface in Ba$_{0.6}$K$_{0.4}$Fe$_2$As$_2$. The inset shows the temperature dependence of the SC gap. (b) Illustration of the gap distribution on 3D Fermi surface in Ba$_{0.6}$K$_{0.4}$Fe$_2$As$_2$. (c) $k_z$ dependence of the symmetrized spectra measured on the $\alpha$ hole Fermi surface in BaFe$_2$(As$_{0.7}$P$_{0.3}$)$_2$. The dashed line is a guide to the eyes for the variation of the SC gap at different $k_z$ values. The inset shows the polarization dependent Fermi surface maps around Z, indicating the $\alpha$ pocket around Z is mainly composed of the $d_z^2$ orbital. (d) Illustration of the gap distribution on the 3D Fermi surface of BaFe$_2$(As$_{0.7}$P$_{0.3}$)$_2$. (e) and (f) The in-plane SC gap distribution on LiFeAs and NaFe$_{1-x}$Co$_{x}$As,  respectively. Data are taken from Refs.~\onlinecite{HDing1,Zhangnode,HDingLiFeAs} and Ref.~\onlinecite{Ge}.} \label{scgap}
\end{figure*}

\begin{figure}[]
\includegraphics[width=8.7cm]{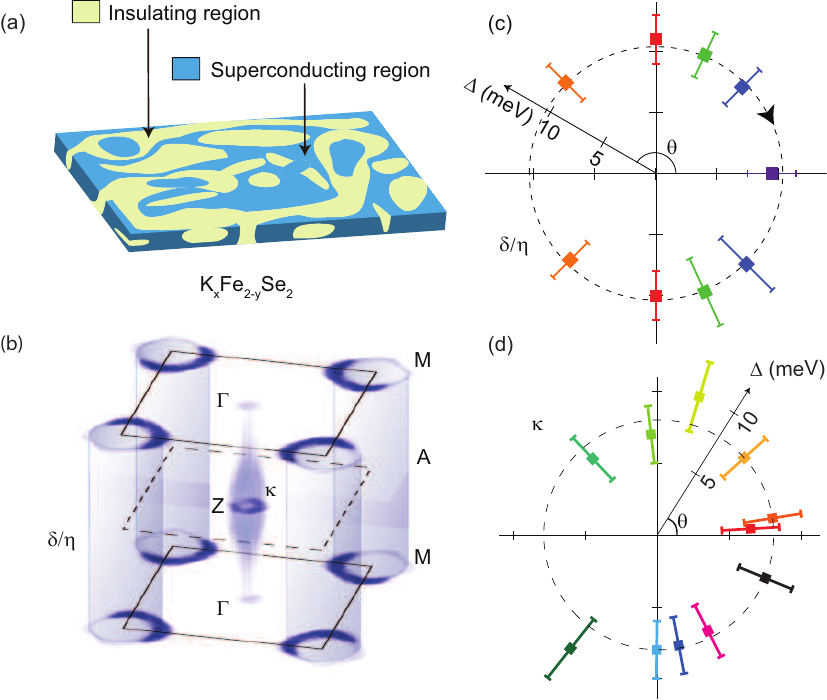}
\caption{(a) Cartoon for mesoscopic phase separation in SC K$_x$Fe$_{2-y}$Se$_2$. (b) The Fermi surface of the SC phase.  (c) Gap distribution on the $\delta$/$\eta$ electron pocket around M in polar coordinates, where the radius represents the gap, and the polar angle $\theta$ represents the position on the $\delta$/$\eta$ pocket with respect to M, with $\theta$~=~0 being the M-$\Gamma$ direction. (d) is the same as (c), but  for the $\kappa$ pocket. Data are taken from Refs.~\onlinecite{KFeSe,ChenPRX} and Ref.~\onlinecite{Xu}.} \label{2nd}
\end{figure}

\begin{figure}[]
\includegraphics[width=8.7cm]{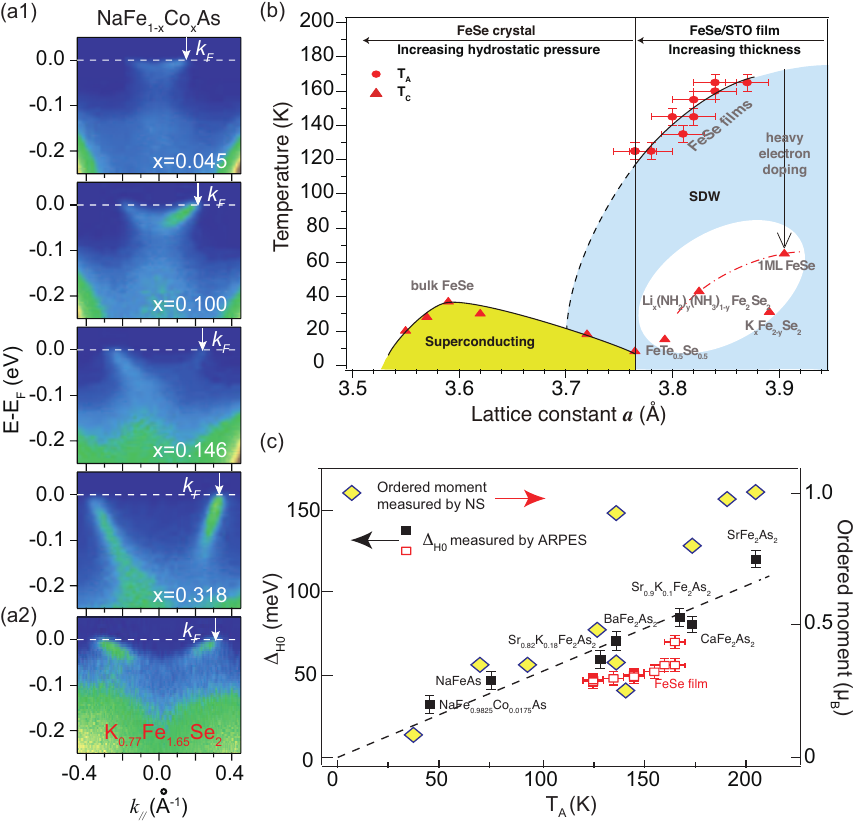}
\caption{(a1) The doping dependence of an electron-like band $\eta$ around the
zone corner for NaFe$_{1-x}$Co$_{x}$As with x=0.045, 0.065, 0.146, and 0.318, respectively. (a2) The electron bands around the zone corner of K$_{0.77}$Fe$_{1.65}$Se$_{2}$. Note that the photon energies used for different samples are not the same but all correspond to the same $k_z$ in the 3D Brillouin zone. (b) Phase diagram of FeSe. The $T_c$ and $T_A$ for FeSe are plotted against the lattice constants. The right side is based on thin film ARPES data, and the left side is based on the transport data of FeSe single crystal under hydrostatic pressure taken from Ref. ~\onlinecite{FeSePressure}. The dashed line represents the extrapolated $T_A$'s, which is the temperature when the band  starts to reconstruct, or become nematic (i.e. $T_A=T_s'$).  Values of $T_c$'s for other iron selenides are also plotted in the elliptical region.   (c) Band separation measured by ARPES and ordered moment measured by  neutron scattering as a function of $T_A$. The dashed line is a guide for the eyes. Data in panel c and d are taken from Ref.~\onlinecite{tan}.  } \label{correlation}
\end{figure}

\subsection{The nematic phases}

The nematic phases here refer to the collinear antiferromagnetic (CAF) state, and the orthorhombic phase below the structural transition temperature $T_s$, as illustrated in Fig.~\ref{nematic}(a) \cite{FisherResistivity}.  There are signs of a ferromagnetic orbital ordered phase above $T_s$, which  exhibits  nematicity as well.\cite{Shin_OO,WKu} Such a nematicity can be viewed in the resistivity of detwined sample shown in Fig.~\ref{nematic}(b). STM has found nematic order with  large periods, which has not been observed by bulk measurements \cite{Davis}.

In the beginning,  the nesting between the electron and hole Fermi surface sheets was considered to be the driving force of the CAF state, as it coincide with the (${\pi,0}$) ordering wavevector \cite{jdong}. As a result, sometimes the CAF state was called  spin density wave. However, it was soon found that a good nesting often does not correspond to a CAF state \cite{zirong}. On the other hand,  when the  hybridization gap occurs, it is well below $E_F$ due to crossing with the folded bands. Moreover, the entire bands shifts, instead of just in the vicinity of the crossing \cite{lxyang,Hecheng}.

The evolution of the nematic electronic structure is illustrated with the example of NaFeAs, where the  $T_s$ is above the Neel temperature \cite{NaFeAsPRB}. The sample was detwinned with moderate pressure, so that the electronic structure along different directions are disentangled, starting from a slightly higher temperature $T_s'$ than $T_s$. As shown in Fig.~\ref{nematic}(c), the $\Gamma$-M$_x$ and  $\Gamma$-M$_y$ are not equivalent in the nematic phases. The electronic structure behaves drastically different in these two directions (Fig.~\ref{nematic}(d,e) in the nematic phases). For the $\beta$ band whose dispersions are the same along these two directions  in the tetragonal paramagnetic (PM) phase, its position starts to move in different directions  below $T_s'$, which is clearly demonstrated by  Fig.~\ref{nematic}(f-h).  The band position difference ($\Delta_{H0}$) saturates at  low temperatures eventually.
 Similar behavior has been observed in detwinned BaFe$_{2-x}$Co$_x$As$_2$ (Fig.~\ref{nematic}(i)) as well, showing that it generally occurs in different compounds \cite{MYi}.
Such a smooth temperature evolution  across   both the structural and magnetic transitions indicates that   they are of the same origin, and  the nematic phases are characterized by the  the same electronic structure nematicity \cite{Hecheng}. Different phases
could be viewed as different stages of the same evolution. At high temperatures, although the structure is tetragonal, the electronic nematicity already occurs above $T_s$, the hopping parameters along $a$ and $b$ start to differ.
As a result, the occupations of $d_{xz}$ and $d_{yz}$ orbitals become inequivalent at all Fe sites, which can be viewed as an ferromagnetic orbital order \cite{WKu}, although such a difference could be rather small, just a few percent in the NaFeAs case down to the lowest temperature \cite{NaFeAsPRB}. Short ranged or fluctuating CAF order might have occurred, in associated with the nematicity. It was suggested that the spin order is more two dimensional and more susceptible to fluctuations, so that the Neel temperature is lower than $T_s$ in some cases \cite{JPHu_nematic}.
The electronic (spin, charge, orbital) and structural degrees of freedom all participate into this process, so that it is likely difficult and unnecessary to identify which is the dominating driving force. Nevertheless, the total electronic energy is reduced significantly, which is well beyond the energy due to the structural change.


The  anisotropy of the resistivity in the nematic phase is consistent with the anisotropic electronic structure, however
recent STM measurements showed that the impurity scattering can be rather anisotropic \cite{Davis}. This explains the variations of the resistivity anisotropy in different compounds. Further investigations are needed to clarify this issue.


FeTe is a special parent compound of iron-based superconductors, which exhibits a bicollinear antiferromagnetic order \cite{WBao,TXiang}. Its polaronic electronic structure is rather different from those of iron pnictides, which is consistent with its large local moment though \cite{FeTePRB,PDai}. The magnetic order in FeTe can be explained by the exchange interactions amongst local moments as well.


\subsection{The SC phase}

There are two critical issues in  the electronic structure of the SC state: 1. what is the pairing symmetry; and 2. what determines the $T_c$.

Pairing symmetry is manifested in the SC gap distribution. For conventional phonon mediated $s$-wave superconductors, the gap is nodeless, i.e. the Fermi surface is fully gapped. While for cuprates, there  are nodes (zero gap) along the diagonal directions, reflecting its $d$-wave symmetry. However, for iron-based superconductors, there are both nodal and nodeless members \cite{HDing1,Terashima,HDing2,linenode1,linenode2,linenode3,linenode4,XQiu,JKDong,QKXue1}.

\subsubsection{Nodeless SC gap}

A large fraction of the iron pnictides are nodeless based on thermal conductivity, penetration depth, STS,  and other measurments \cite{JKDong_BaK,nodeless_FeSe,nodeless_LiFeAs,LDing}. For such a multi-band system, the gap amplitudes vary on different Fermi surface sheets \cite{YanBaK}. On individual Fermi surface sheets, the in-plane gap distribution is often isotropic (within the experimental uncertainty),  eg.  for Ba$_{1-x}$K$_x$Fe$_2$As$_2$ shown in Fig.~\ref{scgap}(a).  The anisotropy along the $k_z$ direction usually is negligible for most Fermi pockets, but some noticeable dependence was found for the $\alpha$ Fermi surface with a sizable warping, as shown in  Fig.~\ref{scgap}(b).
For some compounds, such as LiFeAs and Fe(Te,Se), in-plane gap distribution could be anisotropic as plotted in Fig.~\ref{scgap}(e) , which was attributed to the Fermi surface shape or different pairing interactions mediated by various exchange interaction terms.

Theoretically, the $s^{\pm}$-wave pairing symmetry was proposed  for the iron-based superconductors \cite{KurokiT,Mazin}. However, its proofment requires phase-sensitive techniques to detect the phase difference among various Fermi surface sheets. The magnetic field dependence of the quasiparticle interference pattern of Fe(Te,Se) observed by STM was shown to support the $s^{\pm}$-wave phase-changing scenario \cite{Takagi}.

\subsubsection{Nodal SC gap}

The nodal SC gap was found in some iron pnictides with pnictogen height less than 1.33~\AA, KFe$_2$As$_2$ and FeSe film grown on graphene \cite{linenode4,JKDong,QKXue1}. As shown in Fig.~\ref{scgap}(c-d),  the nodal gap of BaFe$_2$(As$_{1-x}$P$_x$)$_2$ is located in a ring  around Z on the $\alpha$ hole pocket with significant warping and contribution from the $d_{z^2}$ orbital.\cite{Zhangnode}   This indicates its ``accidental" appearance, and rules out the symmetry related origin of the nodes \cite{KurokiT,Mazin}. Theoretically, it has been shown that $d_{z^2}$ does not contribute much to pairing \cite{Kuroki_node}. This hole Fermi pocket is from the same $\alpha$ band whose gap shows significant $k_z$ dependence in the optimally doped Ba$_{1-x}$K$_x$Fe$_2$As$_2$ (Fig.~\ref{scgap}(b)), thus similar origin is expected for both.

For  KFe$_2$As$_2$ and heavily doped Ba$_{1-x}$K$_x$Fe$_2$As$_2$, laser-ARPES work indicate the gap nodes appear at certain points  around Z  on a Fermi pocket with strong $d_{z^2}$  characters \cite{Shin_KFe2As2}, while some other ARPES studies show that they appear on some small hole pockets near M as vertical nodal lines \cite{HDing_BaKnode}.
Whether these are due to sample dependence or due to the high $k_z$ resolution of laser-ARPES needs further clarification. However, the important message is that the nodes in these compounds are accidental, and not due to $d$-wave or other phase changing pairing symmetry. This unites the nodal and nodeless gap behavior in one single scheme.

\subsubsection{Gap in CAF/superconductivity coexisting regime}

As demonstrated by $\mu$SR, neutron scattering and ARPES studies, there is a unique SC regime in the phase diagram of iron-based superconductors, where CAF order and superconductivity coexists \cite{muon_coexistence,NeutronSc&Theory,XHChen_coexistence,Ge,YZhangSrK}. Particularly, the recent STM measurements of NaFe$_{1-x}$Co$_x$As shows the microscopic coexistence and competition of these two orders \cite{Yayu_NaCo}.
The presence of such a unique regime puts strong constraints on the possible paring symmetry.  For example, such a coexistence would not be possible, had the pairing symmetry been $s^{++}$-wave type, where the phases of the SC order parameter are the same on various Fermi surface sheets.
On the other hand, some calculations  based on the $s^{\pm}$-wave pairing suggest that CAF order would induce strong gap anisotropy; and increasing strength of the CAF order,   even nodes could be induced.\cite{theoryMazin,theoryChubukov} Consistently,  a strong gap anisotropy has been observed on the electron Fermi surfaces for NaFe$_{1-x}$Co$_x$As in the coexisting regime, but not in the pure SC regime  (Fig.~\ref{scgap}(f)), although their dopings differ just slightly.

This might explain the nodal gap observed in  the FeSe film grown on a graphene substrate by STS, since  obvious signs of CAF order or strong fluctuations  has been observed by the recent ARPES measurements of multilayer FeSe films grown on STO substrate \cite{QKXue1,tan}.  As a comparison, the STS measurements show that  the gap of Fe(Te,Se)  is nodeless, where  CAF order is not present \cite{Takagi}.

\subsubsection{Electronic features  correlated or uncorrelated with the superconductivity}

$T_c$-determining factors   are crucial for understanding the mechanism of superconductivity in unconventional superconductors. Many  empirical observations have been made as to what affects the $T_c$ for iron-based superconductors.  First of all, it was found  that near the optimal doping of some iron pnictides, certain electron and hole Fermi pockets are better nested, namely, they can overlap on each other  when shifted \cite{Terashima}.  While doped away from the optimal doping, the nesting worsens, since electron and hole pockets change differently. However, various counter examples have been found later.


In some iron pnictides, such as BaFe$_{2-x}$Co$_x$As$_2$ and NaFe$_{1-x}$Co$_x$As, it was found that the superconductivity diminishes when the system is  doped with sufficient electrons so that a Lifshitz transition occurs (specifically, the $d_{xz}/d_{yz}$ based hole pockets disappear) \cite{Kaminski,zirong}. Such a correlation with the superconductivity suggest the importance of this hole Fermi surface. However, later, a counter example is found in Ca$_{10}$(Pt$_4$As$_8$)(Fe$_{2-x}$Pt$_x$As$_2$)$_5$ ($T_c \sim$20 K), where  only $d_{xy}$ based hole pocket exists \cite{XPShen}. A likely cause is that the $d_{xy}$-based bands are strongly scattered by Co dopants in BaFe$_{2-x}$Co$_x$As$_2$ and NaFe$_{1-x}$Co$_x$As, while it is not strongly scattered in  Ca$_{10}$(Pt$_4$As$_8$)(Fe$_{2-x}$Pt$_x$As$_2$)$_5$. This highlights the effects of impurity on the superconductivity. Furthermore, Cr, Mn, Cu and Zn dopants  at the Fe site kill superconductivity much more effectively than Co and Ni. Their effects on the electronic structure have been extensively studied \cite{impurity1,impurity2}.


Electron correlation, or more specifically, spin fluctuations is found to correlate with the superconductivity. It manifests as the distance from the CAF phase, the band renormalization factor, or the dynamical spin susceptibility measured by INS.
These general observations suggest that the superconductivity in iron-based superconductors are mediated by magnetic interactions. In a local pairing scenario, the gap functions of various iron-based superconductors were fitted rather well by including exchange interactions between nearest neighbours, the next  nearest neighbours, and the next next nearest neighbours \cite{Hu}.

So far, a more quantitative correlation between certain electronic properties and superconductivity remains lacking, besides that the SC gap generally scales with  $T_c$. This illustrates the complexity of the problem, and requires further systematic work

\subsection{A$_x$Fe$_{2-y}$Se$_2$ and FeSe thin films}

A$_x$Fe$_{2-y}$Se$_2$ and single layer FeSe  thin film grown on the STO substrate are the two known iron-based superconductors with only electron Fermi surface \cite{KFeSe,tan,QKXue3}. Heavily electron-doped iron pnictides have only electron pocket, however, they are non-SC, as the spin fluctuations diminishes \cite{HDing_KFeAs}. These two chalcogenides with unique electronic structure and rather high $T_c$ pose challenges on the physical pictures established previously for systems with both electron and hole Fermi surface sheets.

\subsubsection{A$_x$Fe$_{2-y}$Se$_2$}

The SC A$_x$Fe$_{2-y}$Se$_2$ sample is phase separated, containing  iron-vacancy ordered insulating domains and SC domains in nanometer scale, as illustrated in Fig.~\ref{2nd}(a).  The electronic structure of the  insulating phase behaves like a Mott insulator.  A semiconducting domain was found in some materials, whose band structure is similar to that of the SC domain, except all the bands are filled. The phase separation was observed both by ARPES and STM, among other measurements \cite{ChenPRX,QKXue2,JQLi}.

For the SC phase, the Fermi surface of K$_x$Fe$_{2-y}$Se$_2$ is shown in Fig.~\ref{2nd}(b). The two electron Fermi surfaces around the zone corner cannot be resolved, and there is a small $\kappa$ electron pocket around Z.\cite{KFeSe} The  gap distributions on these  Fermi surfaces are isotropic as shown Fig.~\ref{2nd}(c)(d) \cite{Xu}. This and the neutron resonance peak \cite{JTPark}, pose severe challenges on theory regarding what kind of pairing symmetry presents in this system.

\subsubsection{FeSe/STO thin films}

The likely high $T_c$ of 65~K in the single layer  FeSe/STO film  has raised a lot of interest \cite{QKXueCPL,QKXue3,tan}. The largest and isotropic SC gap has been obsered by ARPES and STS. The FeSe is found to be doped by electron transferred from the oxygen vacancy states in the STO substrate \cite{tan}. The superconductivity is found only in the first layer on the substrate. For multi-layer FeSe film, which is undoped, there are both electron and hole Fermi surfaces, and the electronic structure reconstruction corresponding to the CAF order is observed \cite{tan}.


By further expanding the FeSe lattice with FeSe/STO/KTO heterostructure, the two electron Fermi pockets  become more elliptical and resolvable \cite{RPeng}. The lack of hybridization between them, and the strong gap anisotropy provide more constraints on theory, suggesting sign change in Fermi sections and interband pair-pair interactions \cite{JPHu_paring}. The gap closes around 70~K in FeSe/STO/KTO, even now 75~K for FeSe/BTO/KTO film \cite{RPengBTO}, indicating a new route to enhance the superconductivity.

\subsection{The critical role of correlations}
Electron correlation is manifested in the band renormalization factor of the iron-based superconductors.   Fig.~\ref{correlation}(a1) illustrates the evolution of dispersion in  NaFe$_{1-x}$Co$_x$As, the band becomes much lighter with increased doping.  In the heavily electron-doped case, the correlation (or spin fluctuation here) is very weak, as they are far away from the CAF phase. As a result, they are non SC. However, its Fermi surface is very similar to that of K$_x$Fe$_{2-y}$Se$_2$, but the latter has a much larger band mass or narrow bandwidth (Fig.~\ref{correlation}(a2)), and a $T_c$ around 30~K. The strong correlation in  K$_x$Fe$_{2-y}$Se$_2$ is likely related to its large lattice constant, since for the multilayer FeSe films, it was found that the CAF ordering strength increases with increased lattice constant  (Fig.~\ref{correlation}(b)). These FeSe films are under high tensile strain, whose lattice constants are much larger than that of bulk FeSe \cite{tan}.

The electronic structure is  itinerant in most of the iron based superconductors. However, there are an effective local moment \cite{PDai}, and the coupling between the itinerant electrons and local moments (which are the two sides of the same coin) gives the Hund's rule coupling, the main correlation source in these compounds \cite{WGYin,ZPYin}. Such a  Hund's metal behavior is also responsible for the CAF order  in these materials. In fact, the energy scale of the electronic structure reconstruction is found to scale with the $T_s$/$T_N$ and  the local moment measured by the neutron scattering Fig.~\ref{correlation}(c) \cite{tan,JJ,neutron1,neutron2,neutron3,neutron4,cxh41,neutron6,neutron7,neutron8,neutron9,neutron10}.

Local moments are important for superconductivity as well. Take systems like the collapsed-tetragonal (cT) phase as an example, where core-electron spectroscopy indicates the absence of local moments, the superconductivity disappears, and bands become less correlated. The electronic structure of the cT phase is similar to that of BaFe$_2$P$_2$, where the small lattice constants  enhance the hopping and thus the itineracy. Relatedly,  for the single layer FeSe/STO film or A$_x$Fe$_{2-y}$Se$_2$, their superconductivity should be related to  the enhanced    correlations   by the expanded lattice.


Assuming the superconductivity of all the iron-based superconductors is due to one unified mechanism, the results obtained on the above two categories of compounds actually  would help to sort out the SC mechanism.
The gap is generally isotropic and nodeless (nodes being accidental) in these systems, although there could be sign changes in different Fermi pockets or sections. The dramatic difference in the Fermi surface topologies  indicates that the pairing is local in the real space, mediated by short range antiferromagnetic interactions.

Finally, we note that although a general experimental phenomenology has been established, there are still many remaining open issues to be addressed. For example, in the recent FeSe/STO studies,  it was suggested that the interface has a non-trivial role on the superconductivity, and particularly the interfacial phonon might play an important role \cite{ZX,RPengBTO}.

\newpage

\section{Theory}

Theory of high-T$_c$ superconductivity remains one of the most fundamental and challenging problems. The BCS theory fails to explain why the SC transition temperatures for both cuprate and iron-based superconductors can be much higher than the possible upper limit for electron-phonon mediated superconductors (40 K) \cite{McMillan}. The highest $T_c \sim 55$ K  of iron-base superconductors~\cite{ZhaoZX2008} is still much lower than that for the cuprate superconductors (164 K under high pressure).  However, the study for the iron-based superconductivity is of particular interest because (1) Fe$^{2+}$ ions have magnetic moments which are generally believed to be detrimental to superconductivity, the discovery of high-T$_c$ superconductivity in iron pnictides has overturned this viewpoint and opens a new direction for exploring new superconductors; (2) there are strong AF fluctuations in iron-based superconductors and the investigation to these materials may help us to understand more deeply the pairing mechanism of high-T$_c$ superconductivity in general.

Iron-based superconductors, including iron pnictides and iron chalcogenites, are quasi-two-dimensional materials. They have very complicated electronic structures and competing interactions. To understand the mechanism of iron-based superconductivity, the first task is to establish the minimal model to describe the low energy electronic excitations in these materials. A key issue under debate is whether the system is in the strong or weak coupling limit, since the interaction that drives electrons to pair can be very different in these two limits.

\subsection{Band structure}
From first principles density functional theory (DFT) calculations,  we know that the low energy excitations of electrons in iron-based superconductors are mainly contributed by Fe 3d electrons. At high temperatures, iron pnictides/chalcogenides are paramagnetic metals. At low temperatures, most of parent compounds, including LaFeAsO,  BaFe$_2$As$_2$  and other 1111 and 122 pnictides, FeTe, are in the AF metallic phase. They become SC upon electron or hole doping.  For LaFePO, LiFeAs or other 111 pnictides, the parent compounds without doping are SC at low temperatures.

\begin{figure}[b]
\begin{center}
\includegraphics[width=0.45\textwidth]{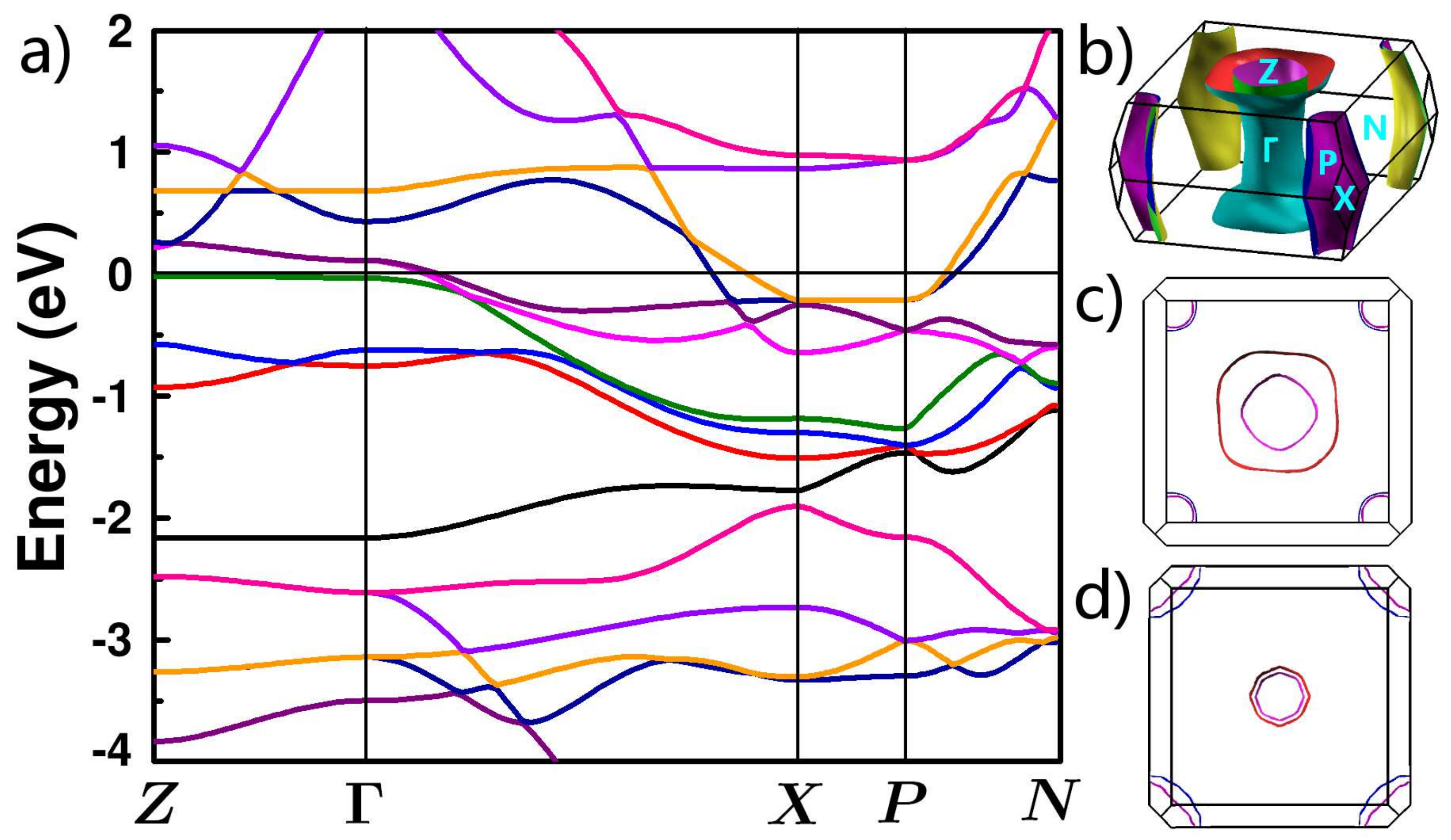}
\caption{(a) Electronic band structure and (b) band structure in the
paramagnetic phase for BaFe$_2$As$_2$   in the folded Brillouin zone.  (c) and (d) are the
sectional views of the Fermi surface through symmetrical k-points
$Z$ and $\Gamma$ perpendicular to the $z$-axis, respectively. (From
Ref.~\cite{Ma122-2008}).} \label{fig:Ba122-band}
\end{center}
\end{figure}

As an example, Fig.~\ref{fig:Ba122-band} shows the Fermi surface and the band structure for BaFe$_2$As$_2$ in the high temperature paramagnetic phase \cite{Ma122-2008}. Similar band structures are found for other iron pnictides \cite{Yildirim2008,Ma2008} and chalcogenides \cite{TXiang}. In general, there are five bands across the Fermi surface. Among them are two electron-like Fermi surfaces centered around $M = (\pi, 0)$ and its equivalent points and two hole-like Fermi surfaces centered around the zone center $\Gamma = (0,0)$. For 1111 or hole doped 122 materials, in addition to the above four surfaces, there is one more hole-like Fermi surface appearing around $Z$. This band is more three-dimensional like than the other four bands and shows a large energy-momentum dispersion along the c-axis.
The band structure of the single-layer FeSe grown on SrTiO$_3$ substrate is relatively simple \cite{Liu2012}. There are only two bands, located around $(\pi, 0)$ and $(0, \pi)$, across the Fermi level \cite{Zhou2012}. The qualitative feature of the band structures obtained by the DFT calculation agrees with the ARPES measurements. But the band width and the effective mass of electrons around the Fermi surface are found to be strongly renormalized by correlation effects which are ignored in the DFT calculation.

\subsection{Minimal model for describing iron-based superconductivity}

Our discussions on minimal model below will be based on strong coupling point of view, where we consider the electron interaction is strong.  The weak coupling viewpoint starts with itinerant electrons and treats electron interaction as a perturbation.  The weak coupling theories may explain a number of experiments in iron-based superconductivity~\cite{Hirschfeld-rev,Chubukov-rev}. However, they have difficult to explain local magnetic moments and superconductivity in systems without Fermi surface nesting.  Among the five 3d orbitals of Fe ions, $d_{xz}$, $d_{yz}$ and $d_{xy}$ contribute most to the low energy excitations. These orbitals couple strongly with each other and with the other two Fe 3d orbitals, $d_{x^2-y^2}$ and $d_{z^2}$, by the Hund's rule exchange interaction. In the atomic limit, an Fe$^{2+}$ ion possesses a large magnetic moment, $\sim 4 \mu_B$,  with a total spin $S = 2$. When Fe atoms form a crystal by hybridizing with As and other atoms, these 3d electrons may become itinerant. If all 3d orbitals of Fe become highly itinerant, one would expect that the magnetic moment of Fe will be completely quenched. However, neutron and other experimental measurements indicate that the magnetic moments of Fe remain finite at least for most of undoped or slightly doped iron pnictides/chalcogenides \cite{cxh41,WBao,Bao2011}.
For example, the ordering moment of Fe in the antiferromagnetic ordered state is about 0.37$\mu_B$ for LaFeAsO~\cite{cxh41}, 2$\mu_B$ for FeTe~\cite{WBao} and 3.31$\mu_B$ for K$_2$Fe$_4$Se$_5$~\cite{Bao2011}. It should be pointed out that the Fermi surface nesting effect cannot give such a large ordering moment and the magnetic moment of Fe must have the contribution from electrons whose energy is well below the Fermi level \cite{Ma122-2008}. The total moment of an Fe ion is a sum of the ordering moment and the fluctuating moment. The fluctuating moment results from thermal and quantum fluctuations of Fe moment and is zero on average. A small ordering moment does not mean the total moment of an Fe spin is also small. The total moments in most of 1111 and 122 pnictides can in fact be much larger than the ordering moments, which suggests strong  quantum fluctuation in the parent compounds.

The existence of Fe moments in these materials means that not all Fe 3d electrons are equally conducting, some of them are more localized than the others. From the first principles density functional calculation, it was found that the crystal splitting of the 3d orbitals is small, but the hybridization between Fe and As/Se atoms and the on-site Coulomb interaction vary differently for different 3d orbitals.  This may lead to an Hund's rule coupling assisted orbital selective Mott transition \cite{Kou2009} and allocate a finite magnetic moment for each Fe by localizing some 3d orbitals. Thus Fe 3d electrons possess both local and itinerant nature. The low energy charge dynamics is governed by itinerant 3d electrons and behaves more like in a conventional metal with weak correlation, whereas the spin dynamics is essentially governed by localized moments and behaves more like in a strong coupling system. Moreover, these itinerant electrons and local moment are not independent, they are actually coupled together by the Hund's rule coupling. This is similar as in a colossal magnetoresistance (CMR) manganate where the double exchange interaction induced by the Hund's coupling between localized and itinerant electrons is important. Of course, the Coulomb screening of conduction electrons to the local moments is stronger in Fe-based materials. This may explain why the magnetoresistance is fairly large in the antiferromagnetic ordered phase in FeTe or other Fe pnicides materials.

Iron-based materials exhibit various antiferromagnetic orders. These orders are driven predominantly by the magnetic interactions between Fe spins, among them the most important one is the superexchange interaction between Fe spins mediated by As or Se 4p electrons \cite{Ma2008}. The superexchange interaction depends on the hybridization between 3d and 4p orbitals, in particular on the bond length and the angle of Fe-As-Fe. Besides this, there is also a direct ferromagnetic exchange interaction between two Fe ions, which is determined by the wavefunction overlap between two 3d orbitals on the two neighboring Fe sites. These exchange interactions are short ranged which extends mainly to the nearest and next nearest neighbors for Fe pnictides or Fe selenides, and to the third next-nearest neighbors for Fe tellurides \cite{Yildirim2008, Ma2008,  Si2008}.

The above discussion suggests that the minimal model for describing iron-based superconductors is approximately given by
\cite{Ma122-2008, Kou2009}
\begin{eqnarray}
H &=& \sum_{ij,\alpha\beta} t_{ij}^{\alpha\beta} c^\dagger_{\alpha, i} c_{\beta, j}
 + J_1\sum_{\langle ij \rangle} S_i \cdot S_j  \nonumber \\
 && + J_2\sum_{\langle\langle ij \rangle\rangle} S_i \cdot S_j ,
\end{eqnarray}
where $\alpha$ and $\beta$ are the orbital quantum number of itinerant electrons, and $S_i = \sum_{\alpha} c^\dagger_{\alpha, i} \mathbf{\sigma} c_{\alpha, i} /2$. The first term is the tight binding Hamiltonian of itinerant electrons.  The second and third terms are the exchange interactions between Fe spins on the nearest and next nearest neighboring sites, respectively. If one ignores the charge fluctuation and considers only the spin dynamics, this Hamiltonian reduces to the $J_1-J_2$ model \cite{Yildirim2008,Ma2008}. In this case, the ground state is collinear AF ordered if $J_2 > J_1/2$. This is indeed the AF order that is observed by neutron scattering measurements \cite{cxh41} in most of the parent compounds of iron-based superconductors.  In passing we note that the minimal model approximates the spin-spin couplings independent of the five d-orbital, and has not included Hund's rule interaction.

In iron-based superconductors, the difference between center momenta for the electron and hole bands, i.e. $M$ and $\Gamma$, coincides with the characteristic wave vector of the $J_2$ term. Thus the $J_2$ term couples strongly with the electron and hole Fermi surfaces. This term is believed to play an important role in driving both AF and SC orders. If the hole and Fermi surfaces are perfectly nested, then the $J_2$ coupling will be strongly amplified in the phase space integration, leading to certain Fermi surface nesting effect, such as the spin density wave instability. Doping can change the phase space that is connected by the nesting vector, which can strengthen SC order and weaken AF order, or vice verse. In general, the competition between SC and AF correlations is strong in these materials. The SC order emerges when the AF order is suppressed.

\subsection{Gap symmetry and structure}

In a SC phase, quantum fluctuations are suppressed by the SC long range order and the BCS mean field approximation is valid. The pairing gap of electrons is an order parameter characterizing a SC state. Physical properties in a SC state can be well described by the BCS theory once the gap function is known. This is the reason why the gap function is of particular interest for study. While a non-s-wave symmetry may indicate unlikely phonon mediated pairing, the pairing symmetry alone is not sufficient to  determined the pairing interaction.

The gap symmetry is determined by the pairing interaction. If the pairing is induced by electron-phonon interaction, it is generally expected that the energy gap has s-wave symmetry.
On the other hand, if the pairing is induced by AF fluctuations,
a spin singlet d-wave (for example in high-T$_c$ cuprates) or spin triplet p-wave (for example in $\mathrm{Sr_2RuO_4}$) pairings are possible, depending strongly on the band structure, especially on the structure of Fermi surfaces. This is because the SC pairing is a low energy effect and involves only excitations of electrons around the Fermi surface. For the same pairing interaction, the gap symmetry may change with the change of the Fermi surface.

The gap symmetry is classified according to the point group of crystal. Theoretical study suggested that the pairing gap of iron-based superconductors has conventional s-wave symmetry \cite{Mazin,Seo2008,Wang-Lee2009,Kuroki2009}, namely in the identity representation of point group.  This has been confirmed by spectroscopy and transport measurements on most of iron-based superconductors including Ba$_{1-x}$K$_x$Fe$_2$As$_2$, BaFe$_{2-x}$Co$_x$As$_2$, KxFe$_{2-x}$Se$_2$, and FeTe$_{1-x}$Se$_x$. However, for KFe$_2$As$_2$, BaFe$_{2-x}$Ru$_x$As$_2$, and nearly all phosphorus-based superconductors, LaFePO, LiFeP, and BaFe$_2$As$_{2-x}$P$_x$, it was found that gap nodes exist. The presence of gap nodes generally implies that the pairing symmetry is unconventional, although an extended s-wave pairing may have accidental nodes on one or more Fermi surfaces.

Fe-base superconductors are multi-band systems. There are several bands across the Fermi level. Even if we assume that the pairing has s-wave symmetry, the relative phases of gap functions can be different on different Fermi surface, depending on inter-band pairing amplitudes to be attractive or repulsive. If the gap function has the same phase on all the Fermi surfaces, the pairing is said to have $s^{++}$ symmetry. On the other hand, if the gap function has opposite phases on different Fermi surfaces, the pairing is said to have $s^{+-}$ symmetry.

The relative phase of the gap function is determined by the interaction between Cooper pairs on different bands. For iron-based superconductors, if the pairing is induced by the AF fluctuations, interaction between Cooper pairs on the electron and hole bands will generally be repulsive. In this case, the SC phases are opposite on the hole and electron Fermi surface, and the gap function has $s^{+-}$ symmetry~\cite{Mazin, Seo2008, Wang-Lee2009, Kuroki2009,ChenWQ}. However, if the pairing is induced by the orbital fluctuation and the SC instability happens in the $A_{1g}$ channel, the interaction between Cooper pairs on the electron and hole bands is attractive, and the gap function will have $s^{++}$ symmetry~\cite{Onari2009}. Thus, from the relative phases in the gap function, one can determine whether the SC pairs are glued by AF fluctuations or by orbital fluctuations.

In the literature, there are quite many discussions on the phase structure of the gap function. However, this phase structure is not in the angular orientation, and is not sensitive to most experiments. It is actually very difficult to resolve unambiguously this seemly simple phase problem~\cite{ChenWQ2}.
For a $s^{+-}$ superconductor, it is expected that a strong neutron resonance peak exists around the momentum linking hole and electron Fermi surfaces, i.e. at $M = (\pi , 0)$ and equivalent points. This resonance peak has in fact been observed in nearly all iron-based superconductors \cite{Inosov2010,LiDai2010}, lending strong support to the theory that predicts the pairing to have $s^{+-}$ symmetry. From the experimental observation of quantum interference of quasiparticles with magnetic or nonmagnetic impurities, it was also found that a $s^{+-}$ pairing is more likely \cite{Takagi}. On the other hand, from Anderson theorem, it is well known that non-magnetic impurity scattering does not affect much on the transition temperature for $s^{++}$ superconductors, but it may reduce strongly the transition temperature for $s^{+-}$ superconductors. In particular, the transition temperature of a $s^{+-}$ superconductor should decrease with increasing impurity concentration. However, for iron-based superconductors, the critical transition temperature does not depend much on the quality of samples. This seems to suggest that the $s^{++}$ pairing is more favored. More systematic study of various impurity effects provides mixed information~\cite{ZheDa}, which may suggest non-universal behavior on the relative phases in iron-based superconductors, in contrast to the universal d-wave pairing in high T$_c$ cuprates.

The SC and AF orders are two competing orders. Generally they repel each other. However, if the pairing has $s^{+-}$ symmetry, theoretical calculation suggested that these two kinds of orders can coexist \cite{Fernandes2010}. Experimentally, this kind of coexistence has indeed been observed in BaFe$_2$As$_2$, Ba$_{1-x}$K$_x$Fe$_2$As$_2$ and SmFeAsO$_{1-x}$F$_x$ with Co substituting Fe or with P substituting As \cite{Laplace2009,AIGoldman5}, and in K$_{x}$Fe$_2$Se$_2$. But in these systems in which the co-existence was observed, the SC gap was also found to have line nodes. It is unknown whether the coexistence is caused by the $s^{+-}$ pairing symmetry or by the line nodes, or the other way around.
\newpage

\section{Summary and Perspective}

In this article, we have reviewed a number of physical properties of iron-based superconductor.  During the past six years, tremendous progress has been achieved in the synthesis of materials, growth of single crystals, characterization of crystal structures, and measurements of thermodynamics, transport and various spectroscopic quantities for iron-based superconductors. This has given us a comprehensive understanding on the chemical and crystal structures, band structures, spin and orbital orderings, pairing symmetry and other physical properties of iron-based superconductors.  In particular, the normal states of Fe-based superconductors have multiple Fermi surfaces including electron Fermi pockets and hole Fermi pockets.  This indicates importance of multi-orbitals in these materials.  Fe-based superconductors are proximate to antiferromagnetism, which suggests that  AF fluctuations are responsible for the observed superconductivity.

 Studies on the mechanism of iron-based superconductivity is an important part of research on the mechanism of high-T$_c$ superconductivity. Any progress in this direction may have strong impact on the study of theory of strongly correlated quantum systems.  To investigate the SC mechanism, one needs to find out the microscopic origin that causes the pairing of electrons and establish a theory that is capable to explain existing experimental data and to predict new experimental effects. This remains a challenging task.  Similar to the cuprate superconductivity, iron-based superconductivity is generally believed to originate pre-dominantly from the electron-electron repulsive interaction, which induces AF fluctuations. Superconductivity induced by AF fluctuation has recently been reviewed by Scalapino~\cite{scalapino}.
 In the present theories based on  AF fluctuation, one approximates the pairing vertex solely in terms of the exchange of AF fluctuations.  This should be reasonable in some cases such as heavy fermion superconductivity, where the SC state is near the AF quantum critical point.  In cuprates, there is also the Mott physics. In iron based SC materials, we have argued that the systems are in strong coupling limit.  Furthermore, the system may well have orbital selected Mott physics. Theoretical description of high Tc superconductivity in both cuprate and iron based superconductivity remains a grant challenge.

Iron based superconductors are multi-band materials. All five 3d orbitals of Fe hybridize strongly with As or Se 4p orbitals. They also couple strongly with each other and have contribution to both itinerant conducting electrons and localized magnetic moments. This brings much complexity to the understanding and explanation of experimental phenomena. We are lacking a clear physical picture with reliable theoretical tools to treat an electronic system with strong coupling between itinerant and localized electrons. Theoretical study for iron-based superconductors relies more on phenomenological analysis of experimental observations and on various approximations.

In short, the iron-based SC mechanism is a challenging problem. To solve this problem, we need to further improve the quality of single crystals and the resolution of measurements. Besides the routine measurements and characterizations, it is more important to design and carry out smoking gun experimental measurements to solve a number of key problems, for example the problem whether the gap function has the $s^{+-}$ symmetry. This will reduce greatly the blindness in the theoretical study and leads to a thorough understanding of iron-based superconductivity.

Fe-based materials have highest SC $T_c$ next to cuprate.  Their discovery has greatly encouraged search for other superconductors  with higher T$_c$.   While we are still far from the stage to predict high T$_c$ materials, there is good progress along this development. It is possible in future that theory may guide the search or synthesis of the high T$_c$ superconductors.

\section{Acknowledgement}
We thank our collaborators in high T$_c$ superconductivity and the colleagues attending Beijing Forum of High-T$_c$ for numerous stimulating discussions over the years.
This work is in part supported by National Science Foundation of China and Ministry of Science and Technology.  PD is also supported by the US NSF DMR-1308603 and OISE-0968226.  XHC, DLF, and FCZ would like to thank Collaborative Innovation Center of Advanced Microstructures, Nanjing, China. We wish to acknowledge Yan Zhou and Xingye Lu for their helpful assistance in the preparation of this manuscript.

\end{document}